\documentclass[
    aps,
    pra,
    notitlepage,
    superscriptaddress,
    longbibliography,
    twocolumn,
    10pt,
]{revtex4-2}

\usepackage{graphicx}
\usepackage{epstopdf}
\usepackage{float}
\usepackage[dvipsnames]{xcolor}

\usepackage{amsmath}
\usepackage{amssymb}
\usepackage{amsfonts}
\usepackage{mathtools}
\usepackage{bm}
\usepackage{physics}

\usepackage{subfigure}
\usepackage{capt-of}
\allowdisplaybreaks

\usepackage{hyperref}
\hypersetup{
    colorlinks = true,
    citecolor  = red,
    linkcolor  = blue,
    urlcolor   = blue,
}

\begin{document}


\title{Quantum Beam-Splitter Cooling and Thermometry in Large Trapped-Ion Crystals}

\author{Kirthik Rajakumar}
\email{kirthik@physics.iitm.ac.in}
\author{Ansh Das}
\author{Abhinay Pandey}
\author{Athreya Shankar}
\email{athreya@physics.iitm.ac.in}

\affiliation{Department of Physics, Indian Institute of Technology Madras,
Chennai 600036, India}
\affiliation{Center for Quantum Information, Communication and Computing, Indian Institute of Technology Madras, Chennai 600036, India}

\date{\today}


\begin{abstract}

We propose and characterize a protocol for rapid near-ground state cooling of the center-of-mass (c.m.) mode of a large trapped ion crystal. When the initial mean thermal occupation of the mode $\bar{n}_i$ is small compared to the number of ions $N$ ($\bar{n}_i \ll N$), a red sideband drive implements a beam-splitter type state exchange (SWAP operation) between the mode and the collective spin of the $N$ ions, with the latter effectively serving as a quantum harmonic oscillator. Subsequently, a reset of the spins removes the entropy, leading to near-ground state cooling of the c.m. mode. We term this protocol as quantum beam-splitter cooling (QBSC). We analyze the impact of several practical imperfections on the final temperature achievable under QBSC, including finite ion number, off-resonant carrier and blue-sideband contributions, the sideband drives arising from spectator modes, and spatially inhomogeneous driving across the crystal. We outline practical strategies to eliminate the carrier drive. We extend our analysis to multiple cycles of QBSC and also generalize the analysis to demonstrate the cooling of non-c.m.\ modes using this protocol. Furthermore, we show that measuring the population statistics of the ions at the end of the SWAP operation can enable near-optimal quantum beam-splitter thermometry (QBST), with the classical Fisher information approaching the quantum Fisher information of a thermal state. We discuss the connection of QBSC with continuous sideband cooling and compare QBST with a recently proposed rapid adiabatic passage-based thermometry scheme. Our work constitutes an example of harnessing many-body effects to open new routes to laser cooling and thermometry in large trapped ion crystals. 
\end{abstract}

\maketitle


\section{Introduction}
\label{sec:intro}

Laser cooling of trapped ions has enabled the exquisite control of single ions and Coulomb crystals of multiple ions for quantum science experiments. Starting from the Doppler cooling limit, sub-Doppler cooling techniques aim to rapidly remove entropy from one or more vibrational modes to bring them close to the motional ground state, in order to improve the fidelity of entangling gates between ions mediated by these modes~\cite{diedrich1989laser,wineland1998experimental,roos2000experimental}. Most theoretical treatments of sub-Doppler laser cooling are confined to a single motional mode of a single ion and to the weak sideband coupling regime, where the coherent electronic-motional coupling strength is weak compared to the effective decay rate of the atom's excited electronic state~\cite{cirac1992laser,morigi2000ground,morigi2003cooling}. In this regime, an effective rate equation can be derived for the cooling of the motion by adiabatically eliminating the internal degrees of freedom of the ions. In contrast, single-ion laser cooling in the strong sideband coupling regime has been explored only recently~\cite{zhang2021fast,khan2026many}. Notably, the strong sideband coupling regime promises a faster cooling rate at the expense of an increased final temperature. 

Beyond single ions, experiments with long ion strings and two-dimensional trapped ion crystals have demonstrated near ground-state cooling of multiple modes of crystals with tens to hundreds of ions~\cite{lechner2016electro,jordan2019near,guo2024asite}. In particular, some of these experiments carried out in the strong sideband coupling regime have observed many-body enhancements in the cooling rate of specific motional modes, suggesting that cooling in the strong sideband coupling regime cannot be treated as independent contributions from individual ions in the crystal~\cite{jordan2019near,shankar2019modeling}. Recent theory work has provided insights into this many-body enhancement, attributing it to an interplay of coherent exchange dynamics between the motional mode and the collective spin of $N$ ions, and spontaneous emission events on the atoms~\cite{khan2026many}. These experimental and theoretical observations raise the question of whether we can deliberately engineer cooling protocols that exploit the large collective spin available in crystals of tens to hundreds of ions to improve the cooling rate and/or reduce the final thermal occupation.

In this paper, we propose a protocol that exploits the collective spin of an $N$ ion crystal to rapidly cool the c.m. mode to near its ground state. Modeling each ion as a spin-1/2 system, we show that when the initial mean thermal occupation of the mode $\bar{n}_i\ll N$, driving the red sideband on resonance leads to an effective quantum beam-splitter (QBS) type interaction between the mode and the collective spin, with the latter effectively acting as a quantum harmonic oscillator. The QBS unitary leads to an effective SWAP operation between the motional state and the spin at a specific time, transferring all the motional excitations to the spin and leaving the mode near its ground state. A subsequent reset of the spins removes the entropy and thereby leads to cooling. We term this protocol as quantum beam-splitter cooling (QBSC). We also show that, if the reset is replaced by a readout of the total population in one of the spin states, the resulting measurements can be used for quantum beam-splitter thermometry (QBST). We demonstrate that the QBST protocol is near-optimal, with the classical Fisher information approaching the quantum Fisher information of a thermal oscillator state in the limit $\bar{n}_i\ll N$. 

This paper is organized as follows. In Sec.~\ref{sec:theory}, we derive the beam-splitter Hamiltonian from the laser-ion crystal interaction. Section~\ref{sec:cooling} analyzes the SWAP-and-reset cooling cycle and characterizes the limiting factors, including finite-size effects, the impact of off-resonant carrier drive and the blue sideband, the role of spectator mode sidebands, and the impact of recoil heating during the reset step. We also discuss strategies to eliminate the carrier contribution. We further examine the effect of a spatially inhomogeneous drive across the crystal, showing that it reduces to the uniform-drive problem for a different collective spin mode and, as a corollary, show that QBSC extends to cooling non-c.m.\ normal modes under a uniform drive. We close Sec.~\ref{sec:cooling} by extending the single-cycle analysis to multiple cycles using a simple cobweb plot that enables a ready inference of the final residual occupation. In Sec.~\ref{sec:thermo}, we analyze the thermometry protocol enabled by spin readout, demonstrate its near-optimality and characterize the limiting factors. In Sec.~\ref{sec:comparisons}, we understand the QBSC protocol in the context of continuous sideband cooling, and compare the QBST protocol against a rapid adiabatic passage (RAP) based thermometry protocol~\cite{lechner2016electro, kirkova2021adiabatic}. We conclude with a summary in Sec.~\ref{sec:conclusion}.


\section{Theory}
\label{sec:theory}



\subsection{The laser-ion crystal Hamiltonian}
\label{subsec:hamiltonian}

We consider $N$ two-level ions of mass $m$ with electronic transition frequency $\omega_0$ confined in a trap. We focus on motion along one spatial direction, say the $z$-axis, and assume that the c.m. vibrational mode frequency in that direction is $\omega$. Denoting the ground (excited) state of each ion by $\ket{0}$ ($\ket{1}$), we consider a traveling wave laser of frequency $\omega_L$ driving the $\ket{0}\leftrightarrow\ket{1}$ transition with carrier Rabi frequency $2\Omega$, and wavevector $\mathbf{k}$ along the $z$ axis. In practice, the effective wavevector can be different from the laser wavevector, e.g., when implementing Raman transitions using a traveling-wave optical lattice, but here we develop the theory in its simplest form. The collective spin is described by $\hat{J}_z = \frac{1}{2}\sum_{l=1}^N\hat{\sigma}_l^z$ and $\hat{J}^{\pm} = \sum_{l=1}^N\hat{\sigma}_l^{\pm}$; the c.m. mode by bosonic operators  $\hat{a}_0$, $\hat{a}_0^{\dagger}$.

In the frame rotating at $\omega_L$, after performing a rotating-wave approximation (RWA) and expanding the laser-ion interaction to first order in the Lamb-Dicke parameter $\eta = k\sqrt{\frac{\hbar}{2m\omega}}$~\cite{wineland1998experimental}, the total laser-ion crystal Hamiltonian is given by ($\hbar=1$ throughout)
\begin{align}
    \hat{H} &= \Delta\hat{J}_z + \omega\hat{a}_0^{\dagger}\hat{a}_0
               + \Omega\!\left(\hat{J}^{+}+\hat{J}^{-}\right) \nonumber \\
            &\quad + \frac{i\eta\Omega}{\sqrt{N}}\!
                 \left(\hat{J}^{+}\hat{a}_0 - \hat{a}_0^{\dagger}\hat{J}^{-}\right) \nonumber \\
            &\quad + \frac{i\eta\Omega}{\sqrt{N}}\!
                 \left(\hat{J}^{+}\hat{a}_0^{\dagger} - \hat{a}_0\hat{J}^{-}\right),
    \label{eq:full_ham}
\end{align}
where $\Delta=\omega_0-\omega_L$ is the ion-laser detuning. The terms in order of appearance in Eq.~(\ref{eq:full_ham}) correspond to the spin energy in the rotating frame, c.m. mode energy, carrier driving (spin flip with no motional coupling) and the two sideband couplings: The term with $\hat{J}^{+}\hat{a}_0$ creates a spin excitation while removing a phonon and vice versa, and is the Jaynes-Cummings (JC) interaction, i.e. the red sideband (RSB) or motion-removing sideband, whereas the term with $\hat{J}^{+}\hat{a}_0^\dagger$ creates both a spin excitation and a phonon, and is the anti-JC interaction, i.e. the blue sideband (BSB) or motion-adding sideband. Both sideband terms in Eq.~(\ref{eq:full_ham}) scale as $\eta\Omega/\sqrt{N}$ since the c.m. mode couples to all $N$ ions uniformly and its single-ion coupling scales down as $1/\sqrt{N}$. 


\subsection{Holstein-Primakoff approximation of the collective spin}
\label{subsec:hp}

To bring out the beam-splitter equivalence of the RSB term, we apply the Holstein-Primakoff (HP) transformation~\cite{holstein1940field, barberena2024fast}, which maps the SU(2) algebra of the collective spin onto a single bosonic mode $\hat{a}_1$, via the exact relations

\begin{equation}
    \hat{J}^{+} = \sqrt{N}\,\hat{a}_1^{\dagger}
    \sqrt{1-\frac{\hat{n}_1}{N}}, \qquad
    \hat{J}^{-} = \sqrt{N}\sqrt{1-\frac{\hat{n}_1}{N}}\,\hat{a}_1,
    \label{eq:hp_exact}
\end{equation}
where $\hat{n}_1 \equiv \hat{a}_1^{\dagger}\hat{a}_1$.  These relations are valid for all $n_1 \leq N$.  In the low-excitation limit $\langle\hat{n}_1\rangle \ll N$, Eq.~\eqref{eq:hp_exact} can be approximated as
\begin{equation}
    \hat{J}^{+} \approx \sqrt{N}\,\hat{a}_1^{\dagger}, \qquad
    \hat{J}^{-} \approx \sqrt{N}\,\hat{a}_1.
    \label{eq:hp0}
\end{equation}
This step constitutes the Holstein-Primakoff approximation (HPA), which shows the approximate equivalence of the collective spin to a harmonic oscillator in the low-excitation limit.


\subsection{Isolating the beam-splitter interaction}
\label{subsec:bsinteraction}

Tuning the laser to the RSB resonance by setting $\Delta = \omega$, we can neglect the carrier and BSB terms as off-resonant drives provided  $\Omega \ll \omega$. Substituting Eq.~\eqref{eq:hp0} for the collective spin operators in Eq.~(\ref{eq:full_ham}) and subsequently moving into a rotating frame at frequency $\omega$ then yields the effective beam-splitter Hamiltonian
\begin{equation}
    \hat{H}_{\mathrm{BS}} = ig\!\left(\hat{a}_1^{\dagger}\hat{a}_0
                              - \hat{a}_0^{\dagger}\hat{a}_1\right), \qquad
    g = \eta\Omega.
    \label{eq:rsb_ham}
\end{equation}
Time evolution under this Hamiltonian results in the beam-splitter unitary $\hat{U}_{\mathrm{BS}}(t) = \exp[gt(\hat{a}_1^\dagger\hat{a}_0 - \hat{a}_0^\dagger\hat{a}_1)]$. In particular, at time
\begin{equation}
    t_{\mathrm{swap}} = \frac{\pi}{2g} = \frac{\pi}{2\eta\Omega},
    \label{eq:tswap}
\end{equation}
the unitary $\hat{U}_{\mathrm{BS}}(t_{\mathrm{swap}})$ implements a  SWAP gate. This aspect is most obvious in the Heisenberg picture, where the operators $\hat{a}_0$ and $\hat{a}_1$ transform as  
\begin{equation}
    \hat{a}_{0}\;\xrightarrow{\hat{U}_{\mathrm{BS}}(t_{\mathrm{swap}})}\;
    -\hat{a}_1,
    \qquad
    \hat{a}_{1}\;\xrightarrow{\hat{U}_{\mathrm{BS}}(t_{\mathrm{swap}})}\;
    \hat{a}_0.
    \label{eq:swap}
\end{equation}
As a result, the motional state is transferred to the collective spin and vice versa. This observation has previously been used to propose protocols for rapid preparation of spin squeezed states by transferring a squeezed motional state into the collective spin degrees of freedom~\cite{barberena2024fast}.

\subsection{Quantum beam-splitter cooling (QBSC) and thermometry (QBST) protocols}
\label{subsec:protocols}

\begin{figure}
    \centering
    \includegraphics[width=\linewidth]{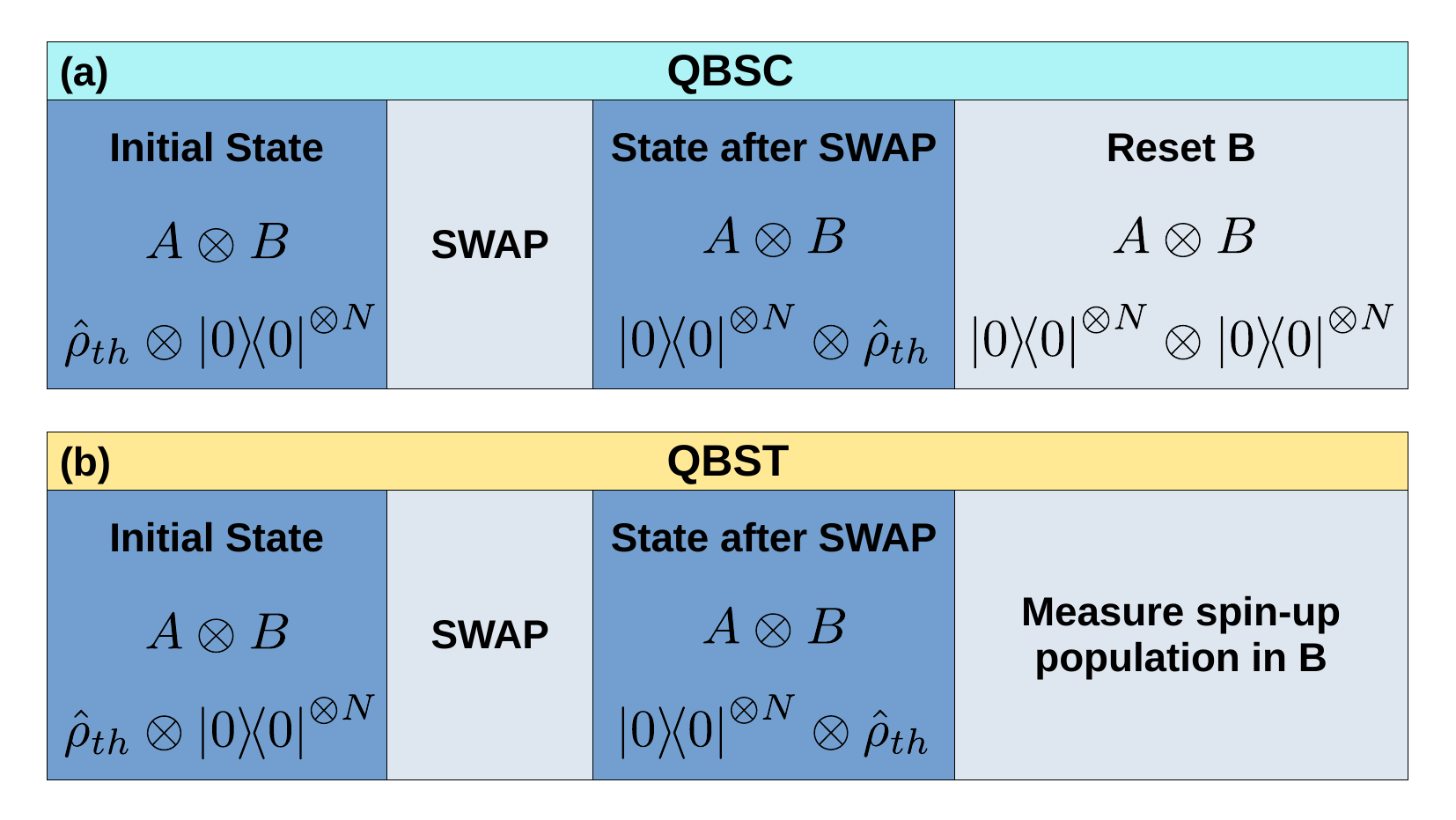}
    \caption{Schematic of the two QBS protocols. Both begin from the product state $\hat{\rho}_{th} \otimes \lvert0\rangle\langle0\rvert^{\otimes N}$, a thermal c.m. phonon distribution with all the ions in the ground state. A resonant red-sideband pulse of duration $t_{\mathrm{swap}}$ [Eq.~(\ref{eq:tswap})] implements the QBS SWAP [Eq.~(\ref{eq:swap})] transferring the phonon distribution to the collective spin-excitation mode. (a) QBSC: An optical-pumping pulse resets all ions to the ground state, removing the transferred entropy irreversibly; iterating this SWAP and reset cools the c.m. mode further towards the ground state. (b) QBST: Reading out the spin-excitation population enables thermometry of $\hat{\rho}_{th}$. }
    \label{fig:protocols}
\end{figure}

The SWAP of Eq.~\eqref{eq:swap} underlies two protocols, one for cooling and the other for thermometry, as follows.

\textit{Quantum beam-splitter cooling (QBSC)}: After an initial Doppler cooling stage followed by a spin reset, the system is assumed to be in the initial product state $\hat{\rho}_{\mathrm{th}}\otimes\lvert 0\rangle\!\langle 0\rvert^{\otimes N}$, where $\hat{\rho}_{\mathrm{th}}$ is  a thermal state of motion with mean occupation $\bar{n}_i$. As shown in Fig.~\ref{fig:protocols}(a), a single cooling cycle consists of two steps: (i) a resonant RSB pulse of duration $t_{\mathrm{swap}}$ that transfers the motional excitations to the collective spin via the SWAP of Eq.~\eqref{eq:swap}; (ii) a fast optical-pumping pulse that resets all ions to their electronic ground state $\lvert 0\rangle$, irreversibly depositing the extracted phonon entropy into the electromagnetic vacuum.  The system is left in the state $\hat{\rho}_{\mathrm{residual}}\otimes\lvert 0\rangle\!\langle 0\rvert^{\otimes N}$ and is ready to be further cooled in a second cycle. Importantly, the SWAP operation is independent of the motional state and enables cooling even if the initial motional state is non-thermal, as may be the case in subsequent cooling cycles. Most of our analysis in Sec.~\ref{sec:cooling} concerns a single cooling cycle, since this isolates the impact of each imperfection cleanly; we return to the effect of iterating the cycle in Sec.~\ref{subsec:multicycle}, where we show that the residual occupation converges to a fixed point under multiple cycles of QBSC.

\textit{Quantum beam-splitter thermometry (QBST)}: Due to near-perfect swapping of spin and motional states in the regime $\bar{n}_i\ll N$, a readout of the number of excited ions after the SWAP operation enables thermometry of the initial motional state, as shown in Fig.~\ref{fig:protocols}(b). We analyze the metrological performance of this protocol in Sec.~\ref{sec:thermo}.



\section{QBSC: Results and Analysis}
\label{sec:cooling}


Unless otherwise stated, the following default parameters for our numerical simulations are: $\eta=0.1$, $\Omega/(2\pi)=0.05$ MHz, $\omega/(2\pi) = 1.6$ MHz and we set $\Delta=\omega$ to resonantly drive the RSB. These parameters lead to the effective RSB drive strength $g/(2\pi)= 5$ kHz, corresponding to $t_{\mathrm{swap}} = 50\,\mu\mathrm{s}$. The value of $\Omega$ is chosen so that $\Omega/\omega=0.031\ll 1$, which ensures that the off-resonant carrier and BSB terms have a negligible impact. 

\subsection{RSB: Impact of finite $N$}
\label{subsec:impact_finite_N}

\begin{figure}[!tb]
    \centering
    \includegraphics[width=\linewidth]{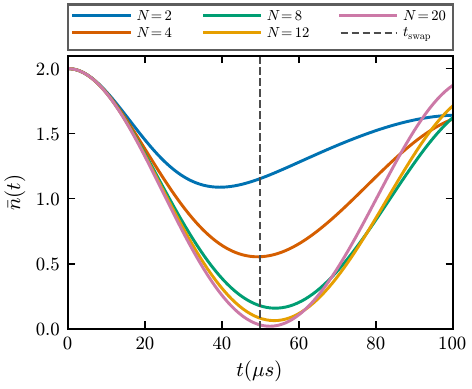}
    \caption{Time evolution of the mean phonon occupation $\bar{n}(t)=\langle\hat{n}_0(t)\rangle$ retaining only the RSB term of Eq.~\eqref{eq:full_ham} for $N=2,4,8,12,20$ ions with $\bar{n}_i=2.0$.  The vertical dashed line marks $t_{\mathrm{swap}}=50\,\mu\mathrm{s}$.}
    \label{fig:evolution}
\end{figure}

We begin our analysis of QBSC by restricting our attention to just the RSB term and investigating the impact of finite $N$. Figure~\ref{fig:evolution} shows the time evolution of the mode occupation $\langle\hat{n}_0(t)\rangle$ under the ideal RSB Hamiltonian for $N=2,4,8,12,20$ ions and $\bar{n}_i=2.0$. The finite $N$ leads to two effects. First, the residual mode occupation at $t=t_{\mathrm{swap}}$ is non-zero. Second, the time at which the minimum occurs is shifted away from $t_{\rm swap}$. However, with increasing $N$, the minimum of $\langle\hat{n}_0(t)\rangle$ approaches $0$ and the time at which the minimum occupation is achieved also shifts closer to $t_{\rm swap}$. 

\begin{figure}[!tb]
    \centering
    \includegraphics[width=\linewidth]{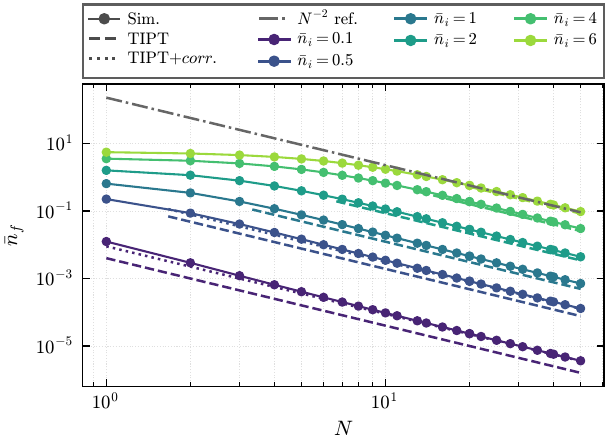}
    \caption{Final mean phonon occupation $\bar{n}_f=\langle\hat{n}_0(t_{\mathrm{swap}})\rangle$ versus $N$ for $\bar{n}_i\in \{0.1,0.5,1.0,2.0,4.0,6.0\}$. The dashed lines show the analytical estimate Eq.~\eqref{eq:residual_eig} obtained from a simplified  time-independent perturbation theory (TIPT) and the dotted lines for $\bar{n}_i=0.1,0.5$ show an improved perturbative estimate (TIPT$+corr.$) in the regime $\bar{n}_i\ll 1$, given by Eq.~\eqref{eq:residual_improved}. The dash-dotted line marks the $N^{-2}$ scaling.}
    \label{fig:residual_loglog}
\end{figure}

To characterize the $N$-dependence, in Fig.~\ref{fig:residual_loglog}, we study the scaling with system size of the residual mode occupation at fixed protocol time $t=t_{\rm swap}$. We plot the final mean occupation $\bar{n}_f$ versus the ion number $N$ for different initial mode occupations $\bar{n}_i$. The figure shows that in the HPA-valid regime $\bar{n}_i/N\ll 1$, $\bar{n}_f$ follows a $1/N^2$ scaling with system size. A simplified first-order perturbation theory calculation (see Appendix~\ref{app:pert}) enables us to obtain analytic expressions that partially explain the finite size correction. We find that the mean residual thermal occupation is given by 
\begin{equation}
\bar{n}_f\approx \frac{\pi^2\bar{n}_i^2(3\bar{n}_i+1)}{32N^2}.
\label{eq:residual_eig}
\end{equation}
The dashed lines in Fig.~\ref{fig:residual_loglog} show the analytical expression, Eq.~(\ref{eq:residual_eig}). The analytical formula reproduces the $N^{-2}$ scaling observed in the simulations. Furthermore, it is also in good quantitative agreement with the numerical final occupations for $\bar{n}_i\gtrsim 1$, where the $\bar{n}_i^3$ term in Eq.~(\ref{eq:residual_eig}) dominates. For $\bar{n}_i \lesssim 1$, the $\bar{n}_i^2$ term dominates, but the analytical result predicts a lower final occupation than the numerical results. We attribute this discrepancy to the fact that our perturbation theoretic calculation only considers diagonal corrections (in a dressed basis) arising from higher-order terms in the HP-transformation but neglects the off-diagonal corrections, which appear to be significant for low $\bar{n}_i$. We explore this in more detail in Appendix~\ref{app:pert}, where we consider the analytically tractable case of exactly $n=2$ initial excitations and find that fully accounting for the perturbation at order $n/N$ yields an additional correction term. In fact, for low initial occupations $\bar{n}_i\ll 1$, the significant correction comes primarily from the $n=2$ sector. Including the correction term leads to the more accurate estimate 
\begin{equation}
\bar{n}_f\approx \frac{\pi^2\bar{n}_i^2(3\bar{n}_i+1)}{32N^2} + \frac{\bar{n}_i^2}{2N^2}.
\label{eq:residual_improved}    
\end{equation}
We plot this refined estimate using dotted lines in Fig.~\ref{fig:residual_loglog} for $\bar{n}_i=0.1$ and $0.5$, and find significantly improved agreement with the numerical results.

\subsection{Off-resonant carrier heating}
\label{subsec:carrier}

In a realistic implementation, the full Hamiltonian Eq.~\eqref{eq:full_ham} retains off-resonant carrier and BSB contributions absent in the pure RSB-based study of the previous section. To characterize their effect, we define three simulation protocols by which terms of Eq.~\eqref{eq:full_ham} are retained alongside the free-energy terms $\Delta\hat{J}_z+\omega\hat{a}_0^\dagger\hat{a}_0$, to correctly capture the cooling dynamics: (i) \textbf{RSB}, keeping only the red-sideband term $\propto i (\hat J^+\hat a_0-\hat a_0^\dagger\hat J^-)$ (the same as the previous section); (ii) \textbf{RSB+BSB}, which adds the blue-sideband term $\propto i (\hat J^+\hat a_0^\dagger-\hat a_0\hat J^-)$; and (iii) \textbf{Carrier-On}, which further adds the carrier term $\Omega(\hat J^+ +\hat J^-)$, i.e.\ the full Hamiltonian Eq.~\eqref{eq:full_ham} with no terms dropped.

\begin{figure}[!tb]
    \centering
    \includegraphics[width = \linewidth]{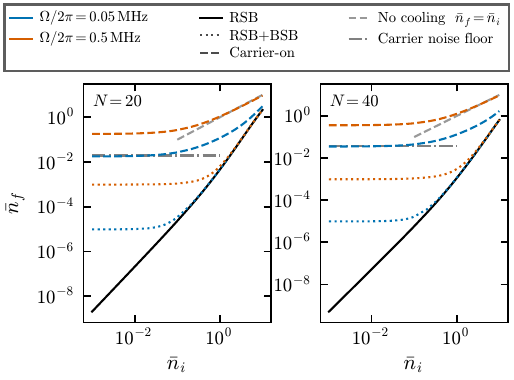}   
    \caption{Final mean phonon occupation $\bar{n}_f$ at $t_{\mathrm{swap}}$ versus initial occupation $\bar{n}_i$ for $N=20$ (left) and $N=40$ (right), comparing RSB (solid black), RSB+BSB (dotted), and Carrier-On (dashed) at $\Omega/2\pi=0.05$ (blue) and $0.5\,\mathrm{MHz}$ (orange) (see Sec.~\ref{subsec:carrier} for the precise Hamiltonian terms retained in each of the three protocols).  The gray dashed diagonal marks no cooling ($\bar{n}_f = \bar{n}_i$). The gray dash-dotted line marks the analytical carrier noise floor [Eq.~(\ref{eq:carrier_floor})], plotted for $\Omega/2\pi = 0.05\,\mathrm{MHz}$ where $\Omega/\omega \ll 1$.}
    \label{fig:carrier_impact}
\end{figure}

Figure~\ref{fig:carrier_impact} shows $\bar{n}_f$ versus $\bar{n}_i$ for $N=20,40$ ions for the three simulation protocols described above. First, we fix $\Omega/2\pi =0.05$ MHz and $\eta=0.1$, and simulate only the RSB and RSB+BSB protocols. We observe that the inclusion of the BSB term has negligible impact on the final temperature for $\bar{n}_i\gtrsim 0.1$ (blue dotted). Further including  the carrier term (`carrier-on' protocol), we find the final occupation to be noticeably higher (dashed blue). For large $\bar{n}_i\gtrsim 2$, the final occupation is lower in the larger crystal with $N=40$. At small $\bar{n}_i \ll 1$, however, the final occupation saturates to a noise floor in contrast to the $N^{-2}$ scaling in the presence of the RSB term alone. In the regime $\Omega/\omega\ll1$, the saturation value can be determined using perturbation theory~(Appendix~\ref{app:carrier_floor}). We find this value to be  
\begin{equation}
    \bar{n}_{\rm floor} = N\left(\frac{\Omega}{\omega}\right)^2
    \label{eq:carrier_floor}
\end{equation}
which is shown by the gray dash-dotted line in both panels of Fig.~\ref{fig:carrier_impact}. Since this floor scales as $N$, larger crystals incur a higher carrier-induced noise floor at small $\bar{n}_i$. As expected, for large amplitude drives, exemplified here with $\Omega/2\pi =0.5$ MHz, we observe hardly any cooling at $t=t_{\rm swap}$ when the carrier is included as shown by the orange dashed lines. However, removing the carrier (RSB+BSB), we find the noise floor due to the BSB is very low even at large drive strengths compared to the respective carrier-on case. These results indicate that QBSC should be implemented with moderate laser amplitudes where the carrier term can be considered to be off-resonant and its impact is negligible. 

As shown by the RSB+BSB curves, eliminating the carrier term can enable the use of larger drive amplitudes that can appreciably speed up QBSC. In practice, two routes can be used to eliminate the carrier term. The first option is to implement QBSC using a standing wave and position the ion crystal at a node. Although laser cooling of ions at standing wave nodes has been achieved~(see e.g. Ref.~\cite{xing2026rapid}), it can be experimentally demanding due to practical challenges with stabilizing the standing wave null with respect to the crystal plane. Furthermore, this method is in general not compatible with 3D crystals. An alternative route is to exploit quantum interference in three-level systems using electromagnetically induced transparency (EIT). Recent work has shown that EIT cooling is effectively equivalent to resolved sideband cooling at the node of a standing wave~\footnote{Up to details in the form of the effective jump operators.}, as the EIT phenomenon cancels the carrier contribution~\cite{khan2026many}. Realizing an EIT configuration in a far-detuned regime, the effective decay rate from the so-called bright state can be suppressed. The coherent coupling induced by the EIT beams is equivalent to the sideband cooling Hamiltonian, Eq.~(\ref{eq:full_ham}), with a suitably dressed dark-bright electronic basis serving as the $\ket{0}$ and $\ket{1}$ levels~\footnote{See Eqs.~(10) and~(11) in Ref.~\cite{khan2026many} for details.}. Therefore, EIT configurations can enable QBSC without a carrier term by implementing the SWAP operation between the motional state and a collective spin defined in this dressed basis. Because this cancellation relies on the internal (dressed-basis) structure of the drive rather than the spatial profile of the field, the EIT approach does not require any ion to occupy a field node, and is therefore a practical option for 3D crystals and also for planar crystals when stabilizing the standing wave node at the crystal plane is challenging.


\subsection{Effect of Spectator Modes}
\label{subsec:spectator}

QBSC targets the cooling of only the c.m. mode, but a physical $N$-ion crystal has $N$ normal modes along any fixed spatial direction. A laser propagating with wavevector $\mathbf{k}$ along the $z$-axis couples to the entire mode branch in that direction. The off-resonant sideband couplings to the $N-1$ spectator modes will in general further degrade the fidelity of the SWAP operation. In this section, we consider a realistic $N$-ion crystal and study the impact of the spectator modes on QBSC. The multi-ion, multi-mode Hamiltonian is given by 

\begin{align}
       \hat{H} &= \frac{\Delta}{2}\sum_{j=1}^N \hat{\sigma}_j^{z}
      + \sum_{\nu=1}^N \omega_\nu \hat{a}_\nu^\dagger \hat{a}_\nu
      + \Omega\sum_{j=1}^N\bigl(\hat{\sigma}_j^{+} + \hat{\sigma}_j^{-}\bigr)
      \nonumber\\
    &\quad
      + i\Omega\sum_{j=1}^N\sum_{\nu=1}^N \eta_{j,\nu}
        \bigl[
          \hat{\sigma}_j^{+}(\hat{a}_\nu + \hat{a}_\nu^\dagger)
          - \hat{\sigma}_j^{-}(\hat{a}_\nu^\dagger + \hat{a}_\nu)
        \bigr].
    \label{eq:H_multimode_main}
\end{align}
Here, $\eta_{j,\nu} = \eta_\nu\,b_{j,\nu}$ with $\eta_\nu$ the Lamb-Dicke (LD) parameter of mode $\nu$ and $b_{j,\nu}$ the amplitude of the mode eigenvector on ion $j$. The LD parameter $\eta_\nu$ can be expressed as 
$\eta_\nu=\eta_1\sqrt{\omega_1/\omega_\nu}$, where we label the c.m. mode as the $\nu=1$ mode. In the following, we set $\Delta=\omega_1$ and study the impact of off-resonant spectator mode couplings on the SWAP operation. We consider an $N=18$ ion chain of $^{40}{\rm Ca}^+$ ions, corresponding to trapping parameters used in Ref.~\cite{lechner2016electro}. We determine the equilibrium configuration and normal modes and consider using QBSC to cool the c.m. mode of one radial branch, with $\omega_1/(2\pi)=2.68~\mathrm{MHz}$. For a radial branch, the c.m. mode is the highest frequency mode and the modes in this case lie in the interval $\omega_\nu/2\pi \in [2.14, 2.68]~\mathrm{MHz}$. For the purposes of demonstration and for simplicity, we do not attempt to match our laser and LD parameters to any specific transition in $^{40}{\rm Ca}^+$ and instead assume our default parameters, i.e. $\Omega/(2\pi)=0.05$ MHz and $\eta=0.1$ as before. We suppose that the modes are initially in thermal equilibrium. We assume the initial thermal occupation of the c.m. mode to be $\bar{n}_{i,1}=2$. From this value, we determine an initial temperature $T_i$ using which we consistently determine the initial thermal occupations of the spectator modes, $\{\bar{n}_{i,\nu}\}$.

Due to the heterogeneous couplings and the multi-mode nature of the Hamiltonian~(\ref{eq:H_multimode_main}), an exact solution for the dynamics is not possible for this system. Instead, we rely on a semiclassical simulation technique based on the truncated Wigner approximation (TWA), which involves sampling the phase space of the initial state according to the Wigner quasiprobability distribution and evolving the phase space variables under classical equations of motion. Due to the hybrid spin-boson structure of the problem, we employ continuous sampling for the bosonic phase space and discrete sampling for the spins~\cite{asier2017noneq}. 

\begin{figure}[!tb]
    \centering
    \includegraphics[width=\linewidth]{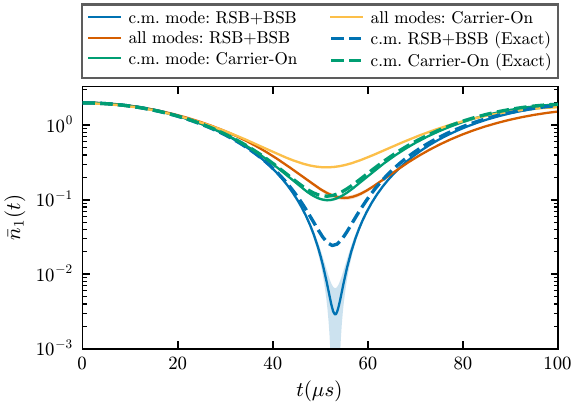}
    \caption{Mean phonon occupation $\bar{n}_1(t)$ of the c.m. mode for the $N=18$ crystal ($\bar{n}_{i,1}=2$, $\eta = 0.1$, $\Omega/2\pi = 50~\mathrm{kHz}$) comparing the time evolution of only the c.m. mode versus all modes. Solid lines show the TWA simulations, while dashed lines provide an exact cross-check for the c.m. mode only configurations.}
    \label{fig:18ions}
\end{figure}

Figure~\ref{fig:18ions} summarizes our study of the impact of spectator modes on QBSC. We first benchmark the performance of the TWA (solid lines) for the c.m. mode-only dynamics, where numerically exact solutions are possible (dashed lines). In the RSB+BSB case, the TWA simulations (blue solid) closely follow the exact dynamics (blue dashed) except in the vicinity of $t=t_{\rm swap}$, where the two disagree by $\Delta\bar{n}\approx0.021$ (exact minimum $\bar{n}\approx0.024$, TWA minimum $\bar{n}\approx0.003$), a relative error of about $88\%$. With the carrier term included (`Carrier-On'), the two methods (green solid and green dashed) disagree by a comparable absolute value, $\Delta\bar{n}\approx0.013$ (exact minimum $\bar{n}\approx0.112$, TWA minimum $\bar{n}\approx0.099$), but here the relative error is only about $12\%$, since the occupation itself stays much higher throughout. Subsequently, we simulate the full multi-mode problem including only the sidebands (RSB+BSB) using the TWA method (orange solid). Spectator modes are observed to have a noticeable impact on the final thermal occupation, elevating the final thermal occupation from approximately $0.02$ to $0.1$, while also noticeably shifting the time corresponding to the minimum occupation. This increase is  approximately at the same level as the increase due to the carrier term alone (green solid/dashed). Finally, including the carrier term and the spectator modes together leads to a further increase in the final thermal occupation at $t=t_{\rm swap}$ by a factor of approximately $3$ to about $0.3$ (yellow solid).

\begin{figure}[!tb]
    \centering
    \includegraphics[width=\linewidth]{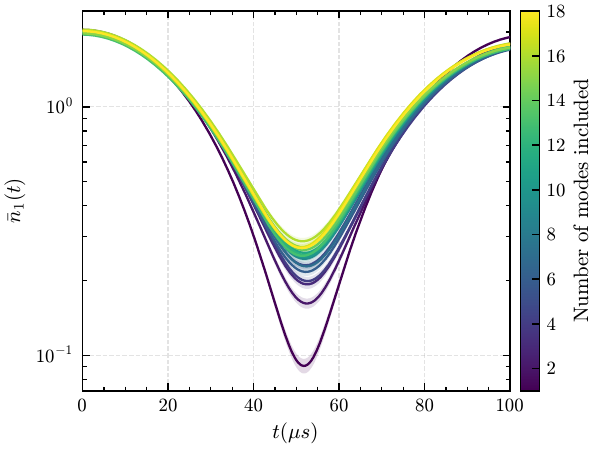}
    \caption{Mean occupation number of the c.m. mode under the Carrier-On protocol, evaluated by sequentially including spectator modes outward from the c.m. mode. Line color indicates the number of modes retained, ranging from $1$ (c.m. mode alone, dark) to $18$ (all modes, light).}
    \label{fig:mode_addition}
\end{figure}

To further unravel the effects of the different spectator modes, we perform TWA simulations of QBSC by progressively including modes farther out from the c.m. mode. The results are shown in Fig.~\ref{fig:mode_addition}. Including only the nearest 3 modes, which are within $35$ kHz of the c.m. mode raises the final thermal occupation at $t=t_{\rm swap}$ from approximately $0.1$ to $0.2$. Further including all the modes progressively raises the final occupation to approximately $0.3$.  These simulations demonstrate the utility of phase space methods in evaluating the detrimental effects of spectator modes on laser cooling protocols. 

\subsection{Recoil heating}

The final reset pulse irreversibly transfers population from the $\ket{1}$ state to the $\ket{0}$ state of the ions. Here, we estimate the recoil heating contribution using a simple model for the optical pumping step. A typical way to reset the spins is to couple $\ket{1}$ to an auxiliary state $\ket{a}$ which decays to $\ket{0}$ with probability $p$ and back to $\ket{1}$ with probability $1-p$. Let us suppose that $n$ of the $N$ ions are in the excited state before the reset pulse. The average number of scattered photons to reset the ions is given by $n/p$. The average energy increase due to a single recoil event is given by the recoil energy $E_r = \hbar^2 k_{\rm sc}^2/2m$, where $k_{\rm sc}$ is the wavevector magnitude for the scattered photon, assumed here to be similar in magnitude for the $\ket{0}\leftrightarrow\ket{a}$ and $\ket{1}\leftrightarrow\ket{a}$ transitions for the purposes of estimation. Assuming isotropic emission for simplicity, the recoil energy is uniformly distributed among the $3N$ motional degrees of freedom of the system. In particular, the energy of the c.m. mode increases on average by $(\Delta E)_{c.m.} = (n/p)E_r/(3N)$. Provided the c.m. mode before the reset pulse is very close to the ground state, the fraction of population in the first Fock state due to recoil heating is then given by $ (\Delta E)_{c.m.}/(\hbar\omega)$. Averaging over the initial thermal mode occupation, the increase in final occupation due to the reset pulse is given by 
\begin{equation}
    \Delta \bar{n}_f = \frac{\bar{n}_i}{p}\frac{1}{3N}\frac{E_r}{\hbar \omega} = \frac{\bar{n}_i\eta_{\rm sc}^2}{3pN}. 
    \label{eq:recoil}
\end{equation}
Here $\eta_{\rm sc}$ is the Lamb-Dicke (LD) parameter associated with the scattering transition, which can in general be different from the LD parameter for the sideband driving. Importantly, the occupation increase due to recoil heating is proportional to the square of the LD parameter and decreases with increasing ion number. Assuming, e.g., $\bar{n}_i=2$, $p=1/2$, $\eta_{\rm sc} = 0.1$, $N=20$, we find $\Delta \bar{n}_f = 7\times 10^{-4}$. In addition, as the $\bar{n}_i$ in subsequent cooling cycles is sharply lowered, the impact of recoil heating becomes negligible with increasing number of cooling cycles (see Sec.~\ref{subsec:multicycle} for the multi-cycle analysis). 

\subsection{Inhomogeneous driving and non-c.m.\ modes}
\label{subsec:inhom}

In practice, the laser intensity can potentially vary over the spatial extent of large crystals. We model this by generalizing the uniform carrier Rabi frequency $\Omega$ of Eq.~\eqref{eq:full_ham} to a site-dependent $\Omega_j$. Mapping each ion onto its own bosonic mode $\hat c_j$ through an individual-ion Holstein-Primakoff approximation (Appendix~\ref{app:inhom}), the carrier and both sidebands all carry the same per-ion weight $\Omega_j$ and therefore combine into the single collective spin mode
\begin{equation}
    \hat{A}_1 \equiv \frac{1}{\tilde\Omega}\sum_{j=1}^N \Omega_j\hat c_j, \qquad \tilde\Omega \equiv \left(\sum_{j=1}^N\Omega_j^2\right)^{1/2}.
    \label{eq:inhom_mode}
\end{equation}
This individual-ion approximation applies the transformation of Sec.~\ref{subsec:hp} to each spin-$1/2$ separately, rather than to the collective spin of magnitude $N/2$. The remaining $N-1$ orthogonal spin modes stay dark, so the Hamiltonian, restricted to the active $(\hat a_0,\hat A_1)$ subspace, takes the form
\begin{align}
    \hat{H}_1 &= \omega\hat a_0^\dagger\hat a_0+\Delta\hat A_1^\dagger\hat A_1+\tilde\Omega\left(\hat A_1^\dagger+\hat A_1\right) \nonumber\\
    &\quad+\frac{i\eta\tilde\Omega}{\sqrt N}\left(\hat A_1^\dagger\hat a_0-\hat a_0^\dagger\hat A_1\right) \nonumber\\
    &\quad+\frac{i\eta\tilde\Omega}{\sqrt N}\left(\hat A_1^\dagger\hat a_0^\dagger-\hat a_0\hat A_1\right).
    \label{eq:inhom_ham}
\end{align}
The individual-ion HPA underlying this reduction is valid provided the excitation of the most strongly driven ion stays small (derived in Appendix~\ref{app:inhom}), i.e.,
\begin{equation}
    \bar n_i\max_j\frac{\Omega_j^2}{\tilde\Omega^2}\ll1,
    \label{eq:inhom_validity}
\end{equation}
which for a uniform profile reduces to $\bar n_i\ll N$ (Sec.~\ref{subsec:hp}). Within this regime, Eq.~\eqref{eq:inhom_ham} is identical in form to Eq.~\eqref{eq:full_ham} after the HPA [Eq.~\eqref{eq:hp0}], under $\Omega\to\tilde\Omega/\sqrt N$ and $\hat a_1\to\hat A_1$. This correspondence gives the swap time,
\begin{equation}
    t_{\rm swap}=\frac{\pi\sqrt N}{2\eta\tilde\Omega}.
    \label{eq:inhom_tswap}
\end{equation}
We assume that the spin reset following the SWAP is unaffected by laser inhomogeneity, e.g., by ensuring a sufficiently long reset pulse. 

\begin{figure}[!tb]
    \centering
    \includegraphics[width=\linewidth]{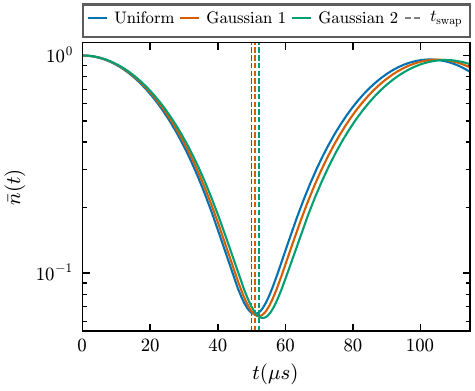}
    \caption{Evolution of the c.m.\ mode under the Carrier-On protocol for the $9$-ion crystal, starting from $\bar n_i=1.0$, comparing a uniform intensity profile (blue) against two Gaussian profiles (1: $90\%$ (orange) and 2: $80\%$ (green) of peak intensity at the outermost ion). Dashed vertical lines mark each profile's $t_{\rm swap}$ [Eq.~\eqref{eq:inhom_tswap}].}
    \label{fig:inhom_verify}
\end{figure}

To test whether the SWAP operation survives an inhomogeneous drive, we simulate the Carrier-On protocol for the c.m.\ mode of one radial branch of a $9$-ion $^{40}$Ca$^+$ crystal, corresponding to trapping parameters used in Ref.~\cite{lechner2016electro}. The c.m.\ mode is the highest-frequency mode of this branch, at $\omega_1/2\pi=2.59$~MHz. We consider three spatial profiles for the laser, that share the same peak Rabi frequency $\Omega_j/2\pi=50.0$~kHz on the central ion: The first profile is uniform and has the same Rabi frequency on every ion, giving $\tilde\Omega/2\pi=150.0$~kHz [Eq.~\eqref{eq:inhom_mode}] and $t_{\rm swap}=50.00\,\mu\mathrm{s}$. The other two are Gaussian profiles that fall to $90\%$ and $80\%$ of the peak intensity at the outermost ion, where $\Omega_j/2\pi=47.4$ and $44.7$~kHz, giving $\tilde\Omega/2\pi=147.0$ and $143.9$~kHz and $t_{\rm swap}=51.02$ and $52.14\,\mu\mathrm{s}$. Evolving all three profiles under the Carrier-On protocol gives the curves of Fig.~\ref{fig:inhom_verify}. The three cooling curves stay close together throughout the cycle, and each profile's minimum occurs close to its own $t_{\rm swap}$ (dashed vertical lines), which shifts as $1/\tilde\Omega$ as expected from Eq.~\eqref{eq:inhom_tswap}. The minimum occupation itself varies by no more than $4.2\%$ across all three profiles, even at the most inhomogeneous profile tested ($80\%$ intensity at the outermost ion). Therefore, moderate coupling inhomogeneities have at most a small effect on the residual phonon number.

Remarkably, the analysis of inhomogeneous driving on the cooling of the c.m.\ mode directly provides insight into the cooling of non-c.m.\ modes under a uniform drive (Appendix~\ref{app:noncm_corres}). Writing ion $j$'s sideband coupling to mode $\nu$ as $\eta_{j,\nu}=\eta_\nu b_{j,\nu}$ [Eq.~\eqref{eq:H_multimode_main}, with $\nu=1$ the c.m.\ mode as in Sec.~\ref{subsec:spectator}] and collecting the ions into $\hat A_\nu\equiv\sum_j b_{j,\nu}\hat c_j$ yields
\begin{align}
    \hat H_\nu = \omega_\nu\hat a_\nu^\dagger\hat a_\nu &+\Delta\hat A_\nu^\dagger\hat A_\nu 
    +i\Omega\eta_\nu\!\left(\hat A_\nu^\dagger\hat a_\nu-\hat a_\nu^\dagger\hat A_\nu\right) \nonumber\\
    &\quad+i\Omega\eta_\nu\!\left(\hat A_\nu^\dagger\hat a_\nu^\dagger-\hat a_\nu\hat A_\nu\right),
    \label{eq:noncm_ham}
\end{align}
which has the structure of Eq.~\eqref{eq:inhom_ham} under $\eta\tilde\Omega/\sqrt N\leftrightarrow\Omega\eta_\nu$. Hence, QBSC cools any normal mode $\nu$, with SWAP time $\pi/(2\Omega\eta_\nu)$ set by that mode's own Lamb-Dicke coupling $\eta_\nu$.

\begin{figure}
    \centering
    \includegraphics[width=\linewidth]{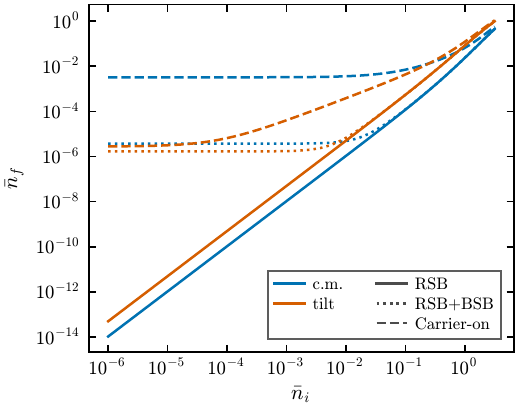}
    \caption{Final mean phonon occupation $\bar n_f$ at $t_{\rm swap}$ versus initial occupation $\bar n_i$, comparing the c.m.\ mode (blue) and the tilt mode (orange) of the same $9$-ion crystal, both driven uniformly, for RSB (solid), RSB+BSB (dotted), and Carrier-On (dashed). The carrier sets a floor for the c.m.\ mode only.}
    \label{fig:noncm_verify}
\end{figure}

Notably, unlike Eq.~\eqref{eq:inhom_ham}, Eq.~\eqref{eq:noncm_ham} contains no carrier term. For a uniform drive, the carrier couples only to $\hat A_1$, the spin mode that is matched to the c.m.\ eigenvector $b_{j,1}=1/\sqrt N$. This mode coincides with $\hat A_\nu$ only for $\nu=1$, reproducing the carrier floor of Sec.~\ref{subsec:carrier}. For $\nu\neq1$, $\hat A_1$ remains uncoupled from the pair $(\hat a_\nu,\hat A_\nu)$, and within the HPA the cooling of a non-c.m.\ mode is insensitive to the carrier term. Figure~\ref{fig:noncm_verify} validates this observation by comparing the c.m.\ mode (blue, $\omega_1/2\pi=2.59$~MHz) with the tilt mode (orange, $\omega_2/2\pi=2.5413$~MHz, the next mode below the c.m.\ mode in the same branch) of the same $9$-ion crystal, both driven uniformly, under the RSB, RSB+BSB, and Carrier-On protocols. We scan $\bar n_i$ over six decades. Under RSB alone (solid), neither curve reaches a floor over the range shown. Under RSB+BSB (dotted), both reach floors of comparable size, $\bar n_f\approx3.7\times10^{-6}$ (c.m.) and $1.7\times10^{-6}$ (tilt), independent of $\bar n_i$ across the range. Under Carrier-On (dashed), the c.m.\ floor is $\bar n_f\approx3.2\times10^{-3}$, likewise independent of $\bar n_i$, whereas the tilt curve keeps decreasing toward its RSB+BSB floor, in line with the absence of a carrier term in Eq.~\eqref{eq:noncm_ham}. At the smallest $\bar n_i=10^{-6}$, it reaches $\bar n_f\approx2.8\times10^{-6}$ and is still decreasing, a factor of $\approx10^3$ below the c.m.\ Carrier-On floor. The carrier is therefore not the limiting factor for the tilt mode; spectator-mode sidebands, excluded from this single-mode treatment, may still set a floor in practice.

\subsection{Multiple cooling cycles}
\label{subsec:multicycle}

While the analysis has so far mostly focused on a single cycle of QBSC, the cooling under multiple cycles can be elegantly interpreted using the single-cycle results. For simplicity, in the multi-cycle analysis, we restrict our attention to the carrier, blue-sideband, and red-sideband terms of the c.m.\ mode. Such a restriction is reasonable, e.g., when the spectator modes are close to their motional ground state. In this case, TWA simulations suggest that the spectator modes have negligible impact on the final occupation of the c.m.\ mode, which is instead limited primarily by the carrier term.

\begin{figure}[!tb]
    \centering
    \includegraphics[width=\linewidth]{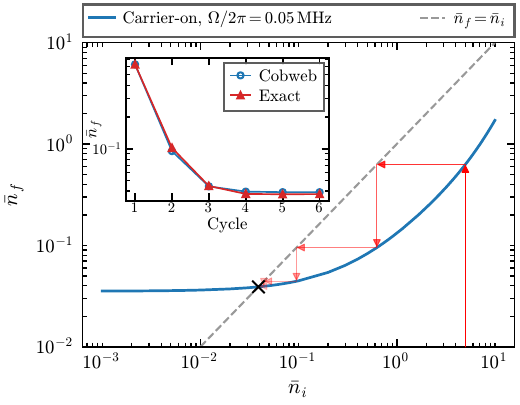}
    \caption{Cobweb construction for the Carrier-On protocol at $N=40$, $\Omega/2\pi=0.05$~MHz. Each cycle's output $\bar n_f$ (read off the curve of Fig.~\ref{fig:carrier_impact}) becomes the next cycle's input, reflected onto the $\bar n_f=\bar n_i$ diagonal; starting from $\bar n_i=5$, the sequence converges to the fixed point marked by the cross. Inset: residual occupation $\bar n_f$ versus cooling cycle for the same case, comparing this cobweb construction (circles) against the exact iterated trajectory (triangles).}
    \label{fig:cobweb}
\end{figure}

Assuming that the final motional state after each cycle is thermal, the curve $\bar n_f(\bar n_i)$ of Fig.~\ref{fig:carrier_impact} is the map of a single QBSC cycle. Feeding each cycle's $\bar n_f$ into the next cycle as $\bar n_i$ gives a cobweb construction for a one-dimensional iterated map. Figure~\ref{fig:cobweb} applies this construction to the Carrier-On protocol for $N=40$ and $\Omega/2\pi=0.05$~MHz. Starting from $\bar n_i=5$, the sequence converges to the fixed point $\bar n_f\approx0.04$ (cross), which equals the carrier floor of Eq.~\eqref{eq:carrier_floor}. The same construction applies to every protocol and drive strength of this section, and the fixed point equals the floor of the respective curve. Repeated cycling therefore does not cool the mode below the single-cycle floor. Recoil heating [Eq.~\eqref{eq:recoil}] scales linearly with $\bar n_i$, so it shrinks along with the cycle and fades out well before either floor is reached. It does not set the multi-cycle limit on its own.

Since this construction assumes the residual state is thermal at the start of every cycle, we check this assumption by separately propagating the exact, non-thermal density matrix through repeated SWAP-and-reset cycles. For the $N=40$, Carrier-On case, shown in the inset of Fig.~\ref{fig:cobweb}, the two trajectories agree almost exactly at the first cycle, since both start from the same genuine thermal state. The second cycle is the first point at which this assumption is actually tested, since the residual state left after the first cycle is not itself thermal. This produces the visible gap there. As subsequent cycles pull both sequences toward their respective fixed points, the gap narrows. It settles at a discrepancy of approximately $4\%$ by the sixth cycle, where $\bar n_f=0.0390$ for the cobweb and $0.0374$ for the exact trajectory, confirming the cobweb construction remains a good approximation even though the actual residual state is not thermal.


\section{QBST: Results and Analysis}
\label{sec:thermo}


Under the leading-order beam-splitter Hamiltonian~\eqref{eq:rsb_ham}, the mean number of spin excitations evolves as
\begin{equation}
    \langle\hat{n}_1(t)\rangle = \bar{n}_i\sin^2(gt).
    \label{eq:therm_signal}
\end{equation}
At $t = t_{\mathrm{swap}}$, the expectation value~\eqref{eq:therm_signal} saturates to $\bar{n}_i$, so a measurement of $\langle\hat{n}_1(t)\rangle$ ideally provides a direct estimate of the initial thermal population and requires no model inversion within the HPA-valid regime $\bar{n}_i \ll N$. However, in practice, finite $N$, the carrier drive and the spectator mode sidebands prevent perfect swapping and a complete estimation model for thermometry including all of these effects can be quite complicated. To illustrate the central ideas, here we restrict our attention to only the c.m. mode and ignore the role of spectator modes. We quantify the metrological performance in terms of the classical Fisher information (CFI) and assess the detrimental role of finite $N$ and off-resonant carrier and BSB drives.

\begin{figure}[!tb]
    \centering
    \includegraphics[width=\linewidth]{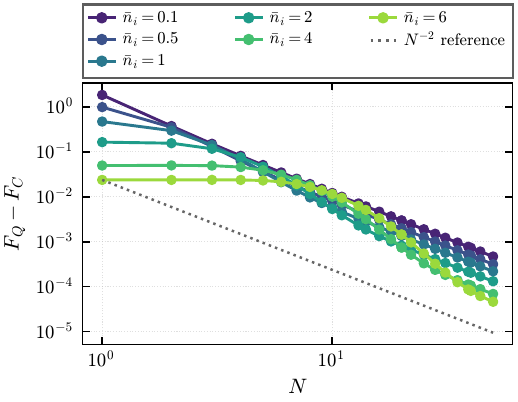}
    \caption{Difference between the quantum Fisher information $F_Q$, Eq.~\eqref{eq:qfi}, and the classical Fisher information $F_C$, Eq.~\eqref{eq:cfi}, versus number of ions $N$ for $\bar{n}_i\in \{0.1,0.5,1.0,2.0,4.0,6.0\}$. The dotted line marks the $N^{-2}$ scaling.}
    \label{fig:qfi-cfi_vs_N}
\end{figure}

We compute the CFI associated with a projective measurement of the number of excitations in the collective spin following the SWAP operation, as shown in Fig.~\ref{fig:protocols}(b). The CFI is given by 
\begin{equation}
 F_C(\bar{n}_i) = \sum_{n_1=0}^{N} [\partial_{\bar{n}_i}\ln P(n_1;\bar{n}_i)]^2 P(n_1;\bar{n}_i),  
 \label{eq:cfi}
\end{equation}
where the index $n_1$ sums over the outcomes of the measurement and $P(n_1;\bar{n}_i)$ is the probability of measuring $n_1$ spins in state $\ket{1}$ given an initial motional mode occupation $\bar{n}_i$. To determine if this measurement is optimal, we compare the CFI to the fundamental quantum Fisher information (QFI) of the thermal state. Since the thermal state is diagonal in the Fock basis, the computation of the QFI is straightforward and leads to 
\begin{equation}
    F_Q(\bar{n}_i) = \frac{1}{\bar{n}_i(\bar{n}_i+1)}.
    \label{eq:qfi}
\end{equation} 

In Fig.~\ref{fig:qfi-cfi_vs_N}, we plot the difference $F_Q-F_C$ as a function of $N$ for different values of $\bar{n}_i$. In the HPA-valid regime where $\bar{n}_i/N\ll 1$, the near-perfect SWAP operation implies that the CFI of the collective spin measurement approaches the QFI of the thermal state as $N$ increases. In this regime, we find that $F_Q-F_C$ scales as $N^{-2}$ with increasing ion number, suggesting an improvement with system size that parallels QBSC.  

\begin{figure}[!tb]
    \centering
    \includegraphics[width=\linewidth]{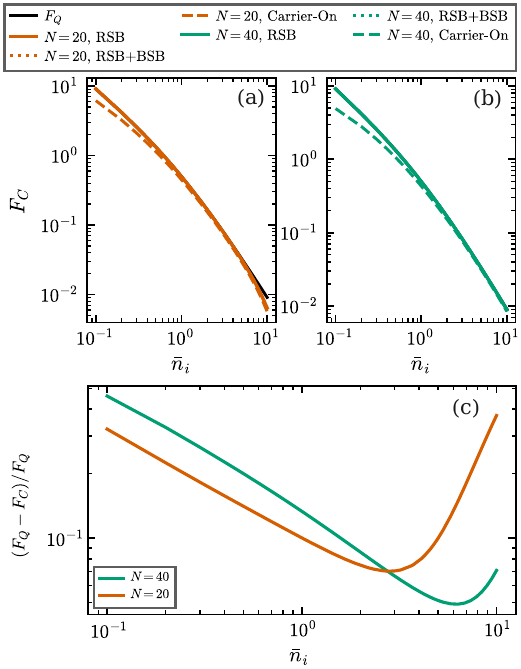}
    \caption{CFI $F_C$ (colored  lines) and QFI $F_Q$ (solid black line) versus $\bar{n}_i$, for (a) $N=20$ (orange), (b) $N=40$ (green) at $\Omega/2\pi=0.05\,\mathrm{MHz}$. Line styles denote the simulation protocols: RSB(Solid), RSB+BSB(dotted), Carrier-On(dashed). (c) Relative fractional difference $(F_Q - F_C)/ F_Q$ versus $\bar{n}_i$ in the Carrier-On case for $N=20$ (orange) and $N=40$ (green).}
    \label{fig:cfi_ions}
\end{figure}

In Fig.~\ref{fig:cfi_ions}, we assess the impact of the BSB and the carrier terms in reducing the CFI for the case of $N=20$ and $40$ ions. We first include the BSB term (dotted lines) and find that it has a negligible impact, similar to the cooling case. The carrier has a noticeable impact on thermometry at low $\bar{n}_i$ values (dashed lines). At intermediate  $\bar{n}_i$ values, the CFI in the presence and the absence of the carrier are nearly identical and close to the QFI, but begin to deviate once again from the QFI with further increase in $\bar{n}_i$, which is evident in the $N=20$ case. This deviation is just the finite $N$ effect. Therefore, in the presence of the carrier term, we can expect an optimal $\bar{n}_i$ where the fractional deviation of the CFI from the QFI is minimal. This expectation is validated by Fig.~\ref{fig:cfi_ions}(c), which plots $(F_Q-F_C)/F_Q$ versus $\bar{n}_i$ for $N=20$ and $N=40$ and confirms the existence of an optimal $\bar{n}_i$ where the fractional deviation is minimized. 

\begin{figure}[!tb]
    \centering
    \includegraphics[width=\linewidth]{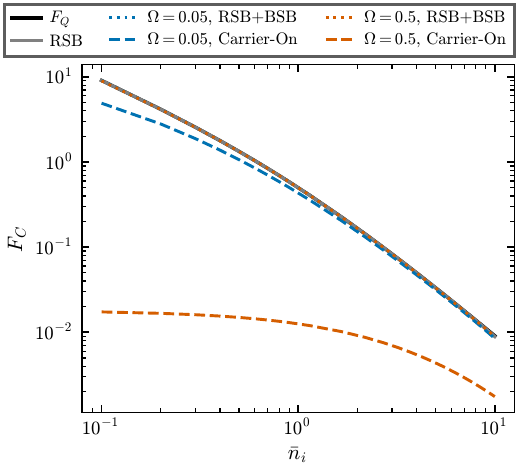}
    \caption{CFI $F_C$ (colored) and QFI $F_Q$ (black solid) versus $\bar{n}_i$ for $N=40$ ions comparing RSB (gray solid), RSB+BSB (dotted), and Carrier-On (dashed) protocols at $\Omega/2\pi=0.05\,\mathrm{MHz}$ (blue) and $0.5\,\mathrm{MHz}$ (orange).}
    \label{fig:cfi_omega}
\end{figure}

An attempt to speed up thermometry by increasing the laser power results in an enhanced carrier contribution that leads to reduced metrological performance. Figure~\ref{fig:cfi_omega} shows the reduction in CFI as $\Omega$ is increased. The simulations indicate that CFI saturates at low $\bar{n}_i$ values for a finite carrier term, as opposed to the QFI which increases as $1/\bar{n}_i$ in this limit. However, removing the carrier and retaining the BSB term, we find that the latter has a negligible effect even at the higher $\Omega$ value.


\section{Discussion}
\label{sec:comparisons}

Having analyzed the performance of the QBSC and QBST protocols, we now discuss them in the context of related cooling and thermometry protocols. 

\subsection{QBSC and sideband cooling}
\label{subsec:sideband_cooling}

In the continuous version of sideband cooling, the RSB is driven resonantly and entropy is removed continuously by an always-on spontaneous-emission channel on the ions that causes the excited state to decay at an effective rate $\gamma$. In contrast, the QBSC protocol of Sec.~\ref{subsec:protocols} separates these two processes in time: A coherent SWAP pulse transfers all phonons to the spin, followed by a fast reset. Furthermore, pulsed sideband cooling, typically modeled in the context of a single ion, is also distinct from QBSC. The former achieves near-ground state cooling in multiple cycles, with each cycle removing exactly one phonon. In contrast, QBSC exploits the collective spin of large trapped-ion crystals and enables removal of multiple phonons in a single cycle. 

Importantly, QBSC can be understood as an extreme limit of continuous sideband cooling in the strong sideband coupling regime, corresponding to the situation when $\gamma\to0$. We make this connection more precise by numerically simulating the master equation for continuous sideband cooling as we tune the system from the weak to the strong sideband coupling regimes. The master equation for sideband cooling is given by

\begin{align}
    \frac{d\hat\rho}{dt} &= -i\left[ \hat{H}, \hat\rho \right ] + \gamma\sum_{l=1}^{N}\mathcal{D}[\hat\sigma_l^{-}]\hat\rho, \label{eq:lindblad_sbc}\\
    & \mathcal{D}[\hat O]\hat\rho \equiv \hat O\hat\rho\hat O^{\dagger}
    - \tfrac{1}{2}\{\hat O^{\dagger}\hat O,\hat\rho\}. \nonumber
\end{align}
Here, we consider $\hat{H}$ from Eq.~(\ref{eq:full_ham}) without the carrier driving term.

\begin{figure}[!tb]
    \centering
    \includegraphics[width=\linewidth]{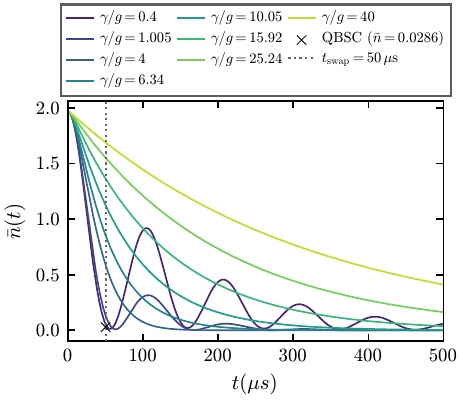}
    \caption{$\bar{n}(t) = \langle\hat n_0(t)\rangle$ for $N=20$, $\bar n_i=2$, evolved under the Carrier-Off Hamiltonian (RSB+BSB) plus the dissipator of Eq.~\eqref{eq:lindblad_sbc} at $\gamma/g\in\{0.4,1,4,6.3,10,16,25,40\}$. The dotted vertical line marks $t_{\mathrm{swap}}=50\,\mu\mathrm{s}$. Cross denotes $\bar{n}(t_{\mathrm{swap}})$ of QBSC obtained using RSB+BSB.}
    \label{fig:sbc_evolution}
\end{figure}

In Fig.~\ref{fig:sbc_evolution}, we plot the cooling curves for an $N=20$ ion crystal as the ratio of $\gamma/g$ is tuned over two orders of magnitude from $40$ (weak sideband) to $0.4$ (strong sideband). Numerically exact simulations for this model are enabled by the permutation symmetry of the master equation~(\ref{eq:lindblad_sbc}), which allows us to use the QuTiP PIQS solver~\cite{shammah2018open}. The figure shows that for large $\gamma/g$ values, the cooling curve follows an exponential decay, consistent with the rate equations typically derived in this regime by adiabatic elimination of the ions' internal states. As $\gamma/g$ is reduced, the system transitions from an overdamped regime to an underdamped regime marked by oscillations in the cooling transient. In the regime $\bar{n}_i\ll N$, a mean-field analysis retaining only the RSB term and the decay channel enables deriving analytical expressions for the cooling rates and reveals that the transition happens at the critical damping value of $\gamma/g=4$~(Appendix \ref{sec:sbc_meanfield}). The QBSC protocol essentially corresponds to the extreme limit $\gamma/g=0$, where the initial half-oscillation corresponds to the SWAP operation of an effective beam-splitter unitary between the motion and the collective spin. The cooling curves demonstrate that at $t=t_{\rm swap}$, QBSC achieves a lower residual occupation than continuous sideband cooling implemented for any non-zero value of the ratio $\gamma/g$.

\subsection{QBST and rapid adiabatic passage}
\label{subsec:adiabatic}

\begin{figure}[!tb]
    \centering
    \includegraphics[width=\linewidth]{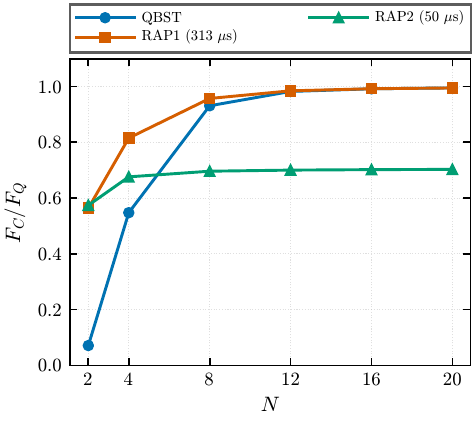}
    \caption{$F_C/F_Q$ versus $N$ for a mean initial occupation of $\bar{n}_i=2$. Data points represent the QBST method (blue circles) and RAP thermometry protocols of two different sweep times: RAP1 ($t=313 \,\mathrm{\mu s}$, orange squares) and RAP2 ($t=50 \,\mathrm{\mu s}$, green triangles).}
    \label{fig:cfi_vs_N}
\end{figure}

A thermometry technique that involves swapping excitations between the mode and the collective spin using rapid adiabatic passage (RAP) has been previously demonstrated and characterized in ion crystals~\cite{lechner2016electro, kirkova2021adiabatic}. Here, we compare QBST against RAP-based thermometry as proposed in Ref.~\cite{kirkova2021adiabatic}. The RAP protocol involves sweeping the laser frequency across the RSB resonance while also suitably varying the coupling strength $g(t)$ over time, according to 
\begin{equation}
    \Delta(t) = \omega +\Delta_0 \sin\!\left(\frac{\mu t}{2}\right), \qquad
    g(t) = g_0 \cos^2\!\left(\frac{\mu t}{2}\right),
    \label{eq:rap_sweep}
\end{equation}
where $\mu$ is the sweep rate and the sweep is performed over the time interval $t\in[-t_{\max},t_{\max}]$ with $t_{\max}=\pi/\mu$. The functional forms ensure that $|\Delta(-t_{\max})-\omega|\gg g(-t_{\max})$ and $|\Delta(t_{\max})-\omega|\gg g(t_{\max})$, which are required for an adiabatic sweep. The coupling amplitude is given by $g_0 = \frac{\eta\Omega}{\sqrt{N}}$ where $\eta = 0.1$ and $\Omega/2\pi = 50\,\mathrm{kHz}$. The detuning amplitude is chosen as $\Delta_0/2\pi = 22\,\mathrm{kHz}$. These parameter choices match the values considered in Ref.~\cite{kirkova2021adiabatic}. Both RAP-based thermometry and QBST have the same objective, which is to map the phonon population on to the ladder of collective spin states, but they achieve it in different ways. RAP does so by adiabatically transferring the population in the Fock state $n$ directly onto a symmetric Dicke state with $n$ excitations. Under perfect adiabatic conditions, this transfer of population is exact as long as $n\leq N$. On the other hand, the SWAP operation in QBST critically relies on the HPA, which is valid only for excitation numbers such that $n \ll N$. The primary advantage of QBST stems from its speed: Adiabatic protocols are typically slow in order to avoid transitions at avoided crossings, whereas QBST does not suffer from this limitation. 

Figure~\ref{fig:cfi_vs_N} summarizes a comparison of the two protocols. We assume $\bar{n}_i=2$ and study the ratio of $F_C/F_Q$ for the two protocols as a function of $N$. In both cases, we neglect off-resonant carrier and BSB terms as well as spectator modes. We assume that the SWAP operation of QBST takes $t_{\rm swap}=50\;\mu$s. For the RAP protocol, we assume two sweep times, one corresponding to $2t_{\rm max}=313 \;\mu$s (RAP1) and a faster sweep with $2t_{\rm max}=50 \;\mu$s (RAP2). The motivation for the choice of RAP1 time is as follows. We fix $N=20$ and scan the sweep rate $\mu$ of the RAP protocol, and determine the fastest $\mu$ (and hence shortest protocol time $2t_{\max}$) for which the CFI of this protocol breaks-even with the CFI of QBST. The choice of RAP2 time is simply to make an equal-time comparison with QBST. For small $N\gtrsim \bar{n}_i$, we observe that RAP1 outperforms QBST, whereas the two protocols have comparable performance in the regime $N\gg \bar{n}_i$. On the other hand, for $N>2$, the faster RAP2 protocol effectively saturates with increasing $N$ and is significantly sub-optimal compared to QBST or RAP1. Therefore, in the HPA-valid regime, QBST provides a faster thermometry protocol while preserving the optimality offered by RAP-based thermometry.

\section{Conclusion}
\label{sec:conclusion}

We have proposed and analyzed quantum beam-splitter cooling (QBSC) and thermometry (QBST), two related protocols that exploit the simplified nature of sideband dynamics in large trapped-ion crystals. In particular, in the limit $\bar{n}_i/N\ll 1$, the collective spin of the ions approximately behaves as a harmonic oscillator and resonant red sideband driving leads to an effective beam splitter interaction that swaps the excitations in the mode into the collective spin. A subsequent reset or readout of the total number of spin excitations enables cooling or thermometry. We have analyzed the role of finite $N$, the carrier and BSB terms, the spectator modes, spatially inhomogeneous driving, and the recoil heating in limiting the final thermal occupation of the target c.m. mode, and find that it can be cooled close to the ground state despite these factors. We further showed that an inhomogeneous drive on the c.m.\ mode is mathematically equivalent to a uniform drive on a different normal mode of the crystal, so QBSC extends beyond the c.m.\ mode to cooling any normal mode of the crystal (Sec.~\ref{subsec:inhom}). Remarkably, unlike the c.m.\ mode, the residual occupation of non-c.m. modes is largely insensitive to the carrier term. Furthermore, the carrier contribution can be altogether eliminated by implementing QBSC and QBST at the node of a standing wave or by exploiting quantum interference in EIT-type configurations, the latter of which remains practical for 2D and 3D crystals since it does not require any ion to occupy a field node. Laser-intensity drift, another practical imperfection, has been addressed by analogous techniques in other trapped-ion contexts, such as active stabilization of the drive power~\cite{king2021,mount2015} or composite-pulse sequences~\cite{mount2015,zlatanov2026}; developing an optimal control protocol tailored to QBSC itself is left for future work. While most of our analysis concerns a single cooling cycle, we also showed, via a cobweb-plot analysis that iterating QBSC converges to a fixed point set by the single-cycle floor. The latter is the analytic floor of Eq.~\eqref{eq:carrier_floor} when the carrier is present, or a much lower RSB+BSB floor once the carrier is eliminated. In the case of thermometry, we demonstrated the near-optimality of the QBST protocol with increasing $N$ by showing that its CFI approaches the QFI of the thermal state. Notably,  this near-optimality persists under moderate carrier drive strength, corresponding to SWAP times of approximately $50 \;\mu$s. 

While continuous ground-state cooling techniques, such as sideband or EIT cooling, aim to cool a part of or the full bandwidth of modes in a chosen spatial direction, QBSC as proposed here is focused on cooling of a single targeted mode. However, the advantage of QBSC stems from its speed; typically, cooling techniques work in the Markovian regime and wait until the motion reaches a steady state. In contrast, QBSC is intrinsically non-Markovian, and enables rapid cooling by splitting the coherent and dissipative processes into two distinct steps. The c.m. mode often has higher heating rates than the other modes~\cite{kalincev2021motional} and hence, QBSC can be used for rapid, targeted cooling of this mode. Such a capability can be useful in platforms such as Penning traps, where the c.m. mode is typically used as the bosonic channel that entangles tens to hundreds of spins for quantum science applications~\cite{bullock2026quantum}.

\emph{Note added:} During the preparation of this paper, we became aware of closely related independent theoretical and experimental work exploring collective enhancement in sideband cooling of ion crystals~\cite{vybornyi2026}.

\section*{Data Availability}
The data that support the findings of this article are openly available~\cite{zenodo_qbs}.

\section*{Acknowledgments}

We thank John Bollinger, Allison Carter, Robert Wolf and Joseph Pham for discussions. We acknowledge support by the Department of Science and Technology, Govt. of India through the INSPIRE Faculty Award (DST/INSPIRE/04/2023/001486), by the Anusandhan National Research Foundation (ANRF), Govt. of India through the Prime Minister's Early Career Research Grant (PMECRG) (ANRF/ECRG/2024/001160/PMS), by IIT Madras through the New Faculty Initiation Grant (NFIG), and the support of the Mphasis F1 Foundation to CQuiCC, IIT Madras.


\appendix

\section{Residual thermal occupation in QBSC: Perturbative analysis}
\label{app:pert}

The Holstein-Primakoff transformation~\cite{holstein1940field} represents the $\mathrm{SU}(2)$ algebra exactly as 
\begin{equation}
\hat{J}^+=\sqrt{N}\,\hat{a}_1^\dagger\sqrt{1-\hat{n}_1/N},
    \qquad
    \hat{J}^-=\sqrt{N}\sqrt{1-\hat{n}_1/N}\,\hat{a}_1,
    \label{eq:hp_exact_app}
\end{equation}
valid for $n_1\leq N$. Expanding to first order in $\hat{n}_1/N$, 
\begin{equation}
    \hat{J}^+\approx\sqrt{N}\hat{a}_1^\dagger-\frac{1}{2\sqrt{N}}\hat{a}_1^\dagger\hat{n}_1,
    \qquad
    \hat{J}^-\approx\sqrt{N}\hat{a}_1-\frac{1}{2\sqrt{N}}\hat{n}_1\hat{a}_1.
    \label{eq:hp_first_order_ops}
\end{equation}
Substituting Eq.~\eqref{eq:hp_first_order_ops} into the RSB term of Eq.~\eqref{eq:full_ham} and collecting order $1/N$ terms gives $\hat{H}_{\mathrm{RSB}}^{(1)}=\hat{H}_0+\hat{V}$, where 
\begin{align}
    &\hat{H}_0 =ig(\hat{a}_1^\dagger\hat{a}_0-\hat{a}_0^\dagger\hat{a}_1),  \nonumber\\
    &\hat{V} =-\frac{ig}{2N}(\hat{a}_1^\dagger\hat{n}_1\hat{a}_0-\hat{a}_0^\dagger\hat{n}_1\hat{a}_1).
    \label{eq:V_app}
\end{align}

Evolving under $\hat{H}_0$ alone, the Heisenberg equations of motion for $\hat{a}_0$ gives us 
$\hat{a}_0(t) = \hat{a}_0(0)\cos(gt) - \hat{a}_1(0)\sin(gt)$, so for the initial state $\hat{\rho}(0) = \hat{\rho}_{\mathrm{th}} \otimes \lvert 0\rangle\!\langle 0 \rvert^{\otimes N}$, the mean phonon occupation is $\langle\hat{n}_0(t)\rangle^{(0)} = \bar{n}_i\cos^2(gt)$, which reaches its first minimum $ \langle\hat{n}_0(t)\rangle^{(0)} = 0 $ exactly at $t_0 = \pi/2g \equiv t_{\mathrm{swap}}$. To estimate the impact of $\hat{V}$, we first diagonalize $\hat{H}_0$ by introducing the polariton operators 
\begin{equation}
    \hat{c}_0 = \frac{1}{\sqrt{2}}(\hat{a}_0 - i\hat{a}_1), \qquad
    \hat{c}_1 = \frac{1}{\sqrt{2}}(\hat{a}_0 + i\hat{a}_1),
    \label{eq:polaritons}
\end{equation}
satisfying  $[\hat{c}_j,\hat{c}_k^\dagger]=\delta_{jk}$. In terms of the  polariton operators, $\hat{H}_0 = g(\hat{n}_{c_0}-\hat{n}_{c_1})$, where $\hat{n}_{c_j} =
\hat{c}_j^\dagger\hat{c}_j$.

Evaluating the diagonal matrix elements $\langle n_0, n_1|\hat{V}|n_0,n_1\rangle$ in the polariton number basis $\ket{n_0,n_1}$, we find the energy shift $\Delta E (n_0,n_1) =  -\frac{g}{4N}(n_0-n_1)(n_0+n_1-1)$. Defining the total excitation $n = n_0 + n_1$, the corrected energy of the state $\ket{n_0,n_1}$ is $
    E = g_n'(n_{c_0} - n_{c_1})$, with the corrected coupling
\begin{equation}
    g_n' = g\left(1-\frac{n-1}{4N} \right).
    \label{eq:gn_prime_app}
\end{equation}
Considering only the above diagonal correction, the oscillations in the $n$-excitation sector occur with a modified frequency $g_n'$ so that for an initial motional Fock state $\ket{n}$~\footnote{ Here, $\ket{n}$ refers to the motional basis states, and not to the polariton basis.}, the mean phonon occupation oscillates at the corrected frequency $\langle\hat{n}_0(t)\rangle_n = n\cos^2(g_n' t)$. 

At $t=t_{\mathrm{swap}}=\pi/2g$, we have 
\begin{equation}
 \cos^2(g_n't_{\mathrm{swap}})=\sin^2\left(\frac{\pi(n-1)}{8N}\right) \approx \left(\frac{\pi(n-1)}{8N}\right)^2   
\end{equation}
for $n\ll N$. Averaging over thermal occupation probabilities leads to the residual thermal occupation 
\begin{equation}
\bar{n}_f = \langle\hat{n}_0(t_{\rm swap})\rangle^{(1)}\approx\sum_n p_n n \left(\frac{\pi(n-1)}{8N}\right)^2.  
\label{eq:nfbar_expression}
\end{equation}
This expression can be evaluated using the thermal factorial moments $\langle n(n-1)\rangle_{\mathrm{th}} = 2\bar{n}_i^2$ and $\langle n(n-1)(n-2)\rangle_{\mathrm{th}} = 6\bar{n}_i^3$. We obtain the result
\begin{equation}
    \bar{n}_f = \left(\frac{\pi}{8N}\right)^2 (6\bar{n}_i^3+2\bar{n}_i^2)  = \frac{\pi^2\bar{n}_i^2(3\bar{n}_i+1)}{32N^2},
\end{equation}
corresponding to Eq.~(\ref{eq:residual_eig}) of the main text.

\subsection{Full perturbative calculation for $n=2$ sector}

Figure~\ref{fig:residual_loglog} shows that while this analytical estimate correctly captures the scaling with $N$ for all values of $\bar{n}_i$, it shows a quantitative discrepancy with the numerical results in the regime $\bar{n}_i \lesssim 1$. For sufficiently small $\bar{n}_i$, the thermal population is primarily concentrated in the $n=0,1$ and $2$ Fock states. The $n=0$ sector is trivial as there are no excitations to exchange. The $n=1$ sector does not show any finite $N$ corrections as a single motional excitation can be perfectly swapped into a finite collective spin at time $t=t_{\rm swap}$. Therefore, $n=2$ is the lowest excitation sector where finite $N$ corrections emerge. Using the Fock basis representation of the bare modes $\hat{a}_0$ and $\hat{a}_1$, this sector is spanned by the three basis states $\{\ket{2,0},\ket{1,1},\ket{0,2}\}$.

The RSB Hamiltonian of Eq.~(\ref{eq:V_app}) can be expressed in the $n=2$ sector as 
\begin{equation}
    H_0 = \sqrt{2}ig M_0, \; \hat{V} = \frac{\sqrt{2}ig}{2N} M_1 
\end{equation}
where $M_0$ and $M_1$ are $3\times 3$ matrices given by  
\begin{equation}
M_0 = \begin{pmatrix}
        0 & -1 & 0 \\
        1 & 0 & -1\\
        0 & 1 & 0
\end{pmatrix}, \;
M_1 = \begin{pmatrix}
        0 & 0 & 0 \\
        0 & 0 & 1\\
        0 & -1 & 0
\end{pmatrix}.
\end{equation}
The unitary matrix governing time evolution until time $t=t_{\rm swap}$ is then given by
\begin{equation}
    {U}(t_{\rm swap}) = e^{A_0+A_1},
\end{equation}
where $A_0 = (\pi/\sqrt{2})M_0$ and $A_1 = [\pi/(2\sqrt{2}N)]M_1$. 

Considering $A_1$ to be a small correction (valid for $N\gg 1$), the approximate form of $U(t_{\rm swap})$ is given by 
\begin{equation}
    U(t_{\rm swap}) \approx e^{A_0} + \int_0^1 dt e^{A_0(1-t)} A_1 e^{A_0 t}. 
\end{equation}
The calculation can be carried out by observing that $M_0$ satisfies $M_0^3=-2M_0$, which leads to a compact form for the exponential:
\begin{equation}
    e^{A_0 t} = I + \frac{\sin(\pi t)}{\sqrt{2}} M_0 + \left(\frac{1-\cos(\pi t)}{2}\right)M_0^2.
\end{equation}
This form can be used to evaluate the explicit $3\times 3$ matrix corresponding to $U(t_{\rm swap})$, which we find to be
\begin{equation}
    U(t_{\rm swap}) \approx \begin{pmatrix}
        0 & 0 & 1\\
        0 & -1 & 0\\
        1 & 0 & 0
    \end{pmatrix}
    +\frac{1}{2N}
    \begin{pmatrix}
        -1 & -\frac{\pi}{2\sqrt{2}}& 0\\
        \frac{\pi}{2\sqrt{2}} & 0 & -\frac{\pi}{2\sqrt{2}}\\
        0 & \frac{\pi}{2\sqrt{2}} & 1
    \end{pmatrix}.
\end{equation}
As a result, for the initial state $\ket{\psi(0)} = \ket{2,0}\equiv (1,0,0)^\dagger$, the state at time $t=t_{\rm swap}$ is given by 
\begin{equation}
    \ket{\psi(t_{\rm swap})} \approx \begin{pmatrix}
        0 \\ 0 \\ 1
    \end{pmatrix}
    + \frac{1}{2N}
    \begin{pmatrix}
        -1 \\
        \frac{\pi}{2\sqrt{2}}\\
        0
    \end{pmatrix}.
\end{equation}
Thus, the residual population at time $t_{\rm swap}$ starting in the $n=2$ motional Fock state is given by 
\begin{equation}
    n_f = \frac{\pi^2}{32N^2} + \frac{1}{2N^2}. 
    \label{eq:n2f_full_perturb}
\end{equation}
Comparing with Eq.~(\ref{eq:nfbar_expression}) where we set $p_2=1$ and $p_n=0$ for all other $n$, we find that the latter formula only predicts the correction term $\pi^2/(32N^2)$ and completely misses the additional correction of $1/(2N^2)$ present in Eq.~(\ref{eq:n2f_full_perturb}). 

For initial thermal states with low $\bar{n}_i\ll 1$, the population in the $n=2$ sector is given by $p_2 \approx \bar{n}_i^2$. Weighting the extra correction term $1/(2N^2)$ by this factor and adding it to Eq.~(\ref{eq:residual_eig}), we obtain the improved analytical estimate Eq.~(\ref{eq:residual_improved}).

\section{Carrier Noise Floor}
\label{app:carrier_floor}

Starting from the full laser-ion Hamiltonian [Eq.~(\ref{eq:full_ham})] with $\Delta = \omega$, after applying the HPA [Eq.~(\ref{eq:hp0})] and dropping the off-resonant BSB term, the Hamiltonian is $\hat{H}_0 + \hat{H}_{\rm car} $, where $\hat{H}_0 = \omega\hat{n}_0 + \omega\hat{n}_1 +  ig\left(\hat{a}_1^\dagger\hat{a}_0 - \hat{a}_0^\dagger\hat{a}_1\right)$ and $\hat{H}_{\rm car} = \Omega\sqrt{N}\left(\hat{a}_1^\dagger+\hat{a}_1\right)$ is the carrier perturbation. Moving to the rotating frame of $\hat{H}_0$, the operators transform as
\begin{align}
    e^{i\hat{H}_0 t}\hat{a}_0\,e^{-i\hat{H}_0 t}
    &= \bigl(\hat{a}_0\cos(gt)-\hat{a}_1\sin(gt)\bigr)e^{-i\omega t}, \nonumber \\
    e^{i\hat{H}_0 t}\hat{a}_1\,e^{-i\hat{H}_0 t}
    &= \bigl(\hat{a}_1\cos(gt)+\hat{a}_0\sin(gt)\bigr)e^{-i\omega t},
    \label{eq:app_modes}
\end{align}
which recover the SWAP of Eq.~(\ref{eq:swap}) at $t_{\rm swap}=\pi/(2g)$. Substituting into $\hat{H}_{\rm car}$, the carrier term in the rotating frame is
\begin{equation}
  \hat{H}_{\rm car}(t)=\Omega\sqrt{N}\Bigl[
    \bigl(\hat{a}_1^\dagger\cos(gt)
         +\hat{a}_0^\dagger\sin(gt)\bigr)e^{i\omega t}
    +{\rm h.c.}\Bigr].
  \label{eq:app_car}
\end{equation}

To calculate the leading contribution due to the carrier, we use first-order time-dependent perturbation theory to find the correction to the state vector. Since the carrier floor is the value to which  $\bar{n}_f$ saturates as $\bar{n}_i\to 0$, we apply perturbation theory starting from the vacuum state. The first-order correction to the state in the rotating frame is
\begin{align}
    |\delta\psi\rangle &= -i\int_0^{t_{\rm swap}}\hat{H}_{\rm car}  (t')|0_0,0_1\rangle dt' \nonumber \\ &= -i\Omega\sqrt{N}\bigl[I_c |0_0,1_1\rangle + I_s |1_0,0_1\rangle \bigr] ,
\end{align}
 where $I_c = \int_0^{t_{\rm swap}}e^{i\omega t'}\cos(gt') dt'$ and $I_s = \int_0^{t_{\rm swap}}e^{i\omega t'}\sin(gt') dt'$. To find the occupation of the motional mode at $t_{\rm swap}$ , we note that, in the rotating frame, $\hat{n}_0$ transforms to $\hat{n}_1$ at the SWAP time. We can then evaluate     
\begin{align}
    \langle \hat{n}_0\rangle_f =  \langle\delta\psi| \hat{n}_1|\delta\psi\rangle = \Omega^2 N\,|I_c|^2.
     \label{eq:app_n0}
\end{align}
Assuming $\omega\gg g$, the integral $I_c$ evaluates to $I_c\approx -1/(i\omega)$. Substituting into Eq.~(\ref{eq:app_n0}), we get 
\begin{equation}
    \bar{n}_{\rm floor}=N\left(\frac{\Omega}{\omega}\right)^2.
\end{equation}

\section{Mean-field analysis of continuous sideband cooling}
\label{sec:sbc_meanfield}

Starting from the master equation~(\ref{eq:lindblad_sbc}) and neglecting the off-resonant BSB term, we write down the equations of motion for the expectation values of the collective dipole $\langle\hat J^-\rangle$ and the mode amplitude $\langle\hat a_0\rangle$. In writing these equations, we assume that the collective spin is close to its ground state throughout the dynamics, which is a valid assumption in the regime $\bar{n}_i/N\ll 1$. This enables us to approximate $\langle\hat J_z\hat a_0\rangle \approx-\frac{N}{2}\langle\hat a_0\rangle$ in the equation for $\langle\hat J^-\rangle$. The equations then reduce to coupled first-order differential equations given by 
\begin{align}
    \frac{d}{dt} \begin{pmatrix} 
    \langle\hat a_0\rangle \\
    \langle\hat J^-\rangle
    \end{pmatrix}
    = \begin{pmatrix} 
    0 & -g/\sqrt{N} \\ 
    g\sqrt{N} & -\gamma/2
    \end{pmatrix} 
    \begin{pmatrix} 
    \langle\hat a_0\rangle \\
    \langle\hat J^-\rangle
    \end{pmatrix}.
    \label{eq:moment_eom}
\end{align}
The decay rates are then given by the eigenvalues of the coupling matrix:
\begin{equation}
    \lambda_\pm = \frac{-\gamma\pm\sqrt{\gamma^2 - 16g^2}}{4}.
    \label{eq:lambda_pm}
\end{equation}

\begin{figure}[!tb]
    \centering\includegraphics[width=\linewidth]{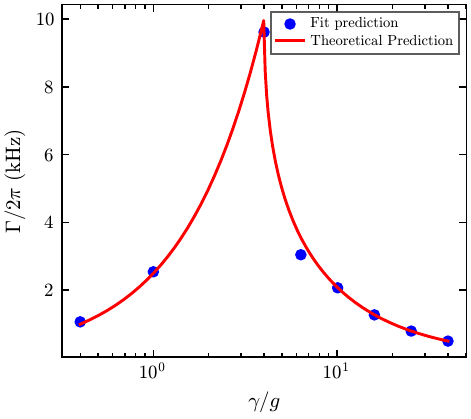}
    \caption{Population decay rate $\Gamma$ from single-exponential fits to the Carrier-Off $\langle\hat n_0(t)\rangle$ of Fig.~\ref{fig:sbc_evolution} (blue circles) and the analytical prediction from Eq.~(\ref{eq:lambda_pm}) (red solid) versus $\gamma/g$.}
    \label{fig:decay_validation}
\end{figure}

The discriminant of Eq.~\eqref{eq:lambda_pm} determines the character of the relaxation, i.e. whether it is underdamped, overdamped or critically damped. The cooling curves in Fig.~\ref{fig:sbc_evolution} exhibit the same behavioral regimes and the corresponding decay rate of the phonon occupation is simply given by twice the amplitude decay rate. Figure~\ref{fig:decay_validation} compares the analytically estimated cooling rate $\Gamma = - 2{\rm Re}\{\lambda_+\}$ with a numerically estimated cooling rate, obtained via fits to the cooling transients. In the underdamped regime $\gamma<4g$, we fit our cooling curves to a function $f(t) = (a+b \cos{(\omega_{osc} t + \phi)})e^{-\Gamma t}+c$. For $\gamma > 4g$, we fit our cooling curves to simple exponential decay $f(t) = a e^{-\Gamma t}+c$. At the critical damping point $\gamma=4g$, since the eigenvalues are degenerate, we construct our fit function to be of the form $f(t) = (a+bt)^2 e^{-\Gamma t}+c$. As seen in Fig.~\ref{fig:decay_validation}, the analytical and numerical rates are in very good agreement, validating the mean-field analysis described here. The maximum cooling rate $\Gamma_{\max} = 2g$ is achieved exactly at critical damping. This is the fastest rate at which the cooling transients approach the steady state in continuous sideband cooling. For $g/2\pi = 5\, \mathrm{kHz}$, this requires $\gamma/2\pi = 20\, \mathrm{kHz}$, which results in $\Gamma_{\max}/(2\pi) = 10$ kHz.

\section{Inhomogeneous laser intensity profile and cooling of non-c.m.\ modes}
\label{app:inhom}

Here we analyze the effect of spatial variation in the laser intensity on the SWAP. Generalizing $\Omega$ of Eq.~\eqref{eq:full_ham} to a site-dependent $\Omega_j$, each ion maps onto its own bosonic mode $\hat c_j$ through the exact Holstein-Primakoff relations $\hat\sigma_j^+=\hat c_j^\dagger\sqrt{1-\hat n_j}$ and $\hat\sigma_j^z=2\hat n_j-1$, with $\hat n_j\equiv\hat c_j^\dagger\hat c_j$~\cite{holstein1940field}. The individual-ion approximation $\sqrt{1-\hat n_j}\approx1$, applied to every ion separately, leaves each driving term proportional to a single power of $\hat c_j$ or $\hat c_j^\dagger$, weighted by $\Omega_j$. Since the carrier and both sidebands carry the same weights, all three reduce to the same collective mode $\hat A_1\equiv\tilde\Omega^{-1}\sum_j\Omega_j\hat c_j$ [Eq.~\eqref{eq:inhom_mode}], via $\sum_j\Omega_j\hat c_j=\tilde\Omega\hat A_1$. Extending $\{\Omega_j/\tilde\Omega\}$ to an orthonormal basis of $\mathbb R^N$ (e.g., by Gram-Schmidt orthogonalization) defines $N-1$ additional collective modes $\hat A_2,\ldots,\hat A_N$, orthogonal to $\hat A_1$ and to each other. These satisfy $\sum_j\hat n_j=\sum_{k=1}^N\hat A_k^\dagger\hat A_k$, and the Hamiltonian becomes
\begin{align}
    \hat H &= \Delta\sum_{k=1}^N\hat A_k^\dagger\hat A_k+\omega\hat a_0^\dagger\hat a_0+\tilde\Omega(\hat A_1^\dagger+\hat A_1) \nonumber\\
    &\quad+\frac{i\eta\tilde\Omega}{\sqrt N}\left(\hat A_1^\dagger\hat a_0-\hat a_0^\dagger\hat A_1\right)+\frac{i\eta\tilde\Omega}{\sqrt N}\left(\hat A_1^\dagger\hat a_0^\dagger-\hat a_0\hat A_1\right).
    \label{eq:inhom_ham_full}
\end{align}
In Eq.~\eqref{eq:inhom_ham_full}, $\hat A_2,\ldots,\hat A_N$ enter only through their number operators, so $[\hat H,\hat A_k^\dagger\hat A_k]=0$ for $k\geq2$. Starting from the ion vacuum $\ket{0}\!\bra{0}^{\otimes N}$, these modes therefore remain unpopulated throughout the dynamics. Retaining only terms with $\hat A_1$, Eq.~\eqref{eq:inhom_ham_full} reduces to Eq.~\eqref{eq:inhom_ham}.

Setting $\Delta=\omega$ and dropping the off-resonant carrier and BSB terms, as in Sec.~\ref{subsec:bsinteraction}, gives the beam-splitter Hamiltonian of Eq.~\eqref{eq:rsb_ham} with
\begin{equation}
    g\to\frac{\eta\tilde\Omega}{\sqrt N}.
    \label{eq:g_sub}
\end{equation}
This Hamiltonian therefore generates the same beam-splitter unitary for $\hat A_1$ as in the uniform-drive case, which implements a SWAP at time $t_{\rm swap}$. Since $\hat A_2,\ldots,\hat A_N$ remain in the vacuum, inverting Eq.~\eqref{eq:inhom_mode} gives the excitation transferred to ion $j$ once the SWAP completes, $\langle\hat n_j\rangle=(\Omega_j/\tilde\Omega)^2\bar n_i$. The individual-ion approximation therefore holds at every site provided the bound of Sec.~\ref{subsec:inhom} is satisfied,
\begin{equation}
    \bar n_i\,\frac{\max_j\Omega_j^2}{\tilde\Omega^2}\ll 1.
\end{equation}
This bound tightens for a profile concentrated on a few ions, since $\max_j\Omega_j^2/\tilde\Omega^2\geq1/N$ for any profile, with equality only for a uniform drive. Within this regime, Eq.~\eqref{eq:g_sub} gives both the SWAP time of Eq.~\eqref{eq:inhom_tswap} and, via the perturbative argument of Appendix~\ref{app:carrier_floor}, the generalized carrier floor $\bar n_{\rm floor}=(\tilde\Omega/\omega)^2$. We have verified this floor numerically. Both results reduce to their uniform-drive values for $\Omega_j=\Omega$, where $\tilde\Omega=\sqrt N\,\Omega$. The finite-$N$ correction of Appendix~\ref{app:pert} and the continuous-cooling analysis of Appendix~\ref{sec:sbc_meanfield} assume a uniform drive and do not extend to this case.

\subsection{Correspondence with non-c.m.\ mode cooling}
\label{app:noncm_corres}

Applying the same steps to a uniform drive ($\Omega_j=\Omega$) on the multi-mode Hamiltonian of Eq.~\eqref{eq:H_multimode_main} defines collective spin modes aligned with the normal-mode eigenvectors of the crystal instead of the drive profile,
\begin{equation}
    \hat A_\mu=\sum_{j=1}^N b_{j,\mu}\hat c_j, \qquad \mu=1,\ldots,N.
    \label{eq:noncm_modes}
\end{equation}
These modes are orthonormal since $\sum_j b_{j,\mu}b_{j,\mu'}=\delta_{\mu\mu'}$. Since $b_{j,1}=1/\sqrt N$ gives $\sum_j\hat c_j=\sqrt N\hat A_1$, the carrier term $\Omega\sum_j\hat\sigma_j^+\approx\Omega\sum_j\hat c_j^\dagger$ reduces, via $\sum_j\hat c_j=\sqrt N\hat A_1$, to a term acting on $\hat A_1$ alone. With the per-ion sideband coupling $\eta_{j,\mu}=\eta_\mu b_{j,\mu}$, the same orthogonality relation decouples the sideband terms mode by mode, so each sideband term couples only to its own $\hat A_\mu$, giving
\begin{align}
    \hat H &= \Delta\sum_{\mu=1}^N\hat A_\mu^\dagger\hat A_\mu+\sum_{\mu=1}^N\omega_\mu\hat a_\mu^\dagger\hat a_\mu+\Omega\sqrt N\left(\hat A_1^\dagger+\hat A_1\right) \nonumber\\
    &\quad+i\Omega\sum_{\mu=1}^N\eta_\mu\left[\hat A_\mu^\dagger\left(\hat a_\mu+\hat a_\mu^\dagger\right)-\hat A_\mu\left(\hat a_\mu^\dagger+\hat a_\mu\right)\right].
    \label{eq:noncm_ham_full}
\end{align}
Equation~\eqref{eq:noncm_ham_full} separates into $N$ independent motional-spin pairs $(\hat a_\mu,\hat A_\mu)$ plus the carrier term on $\hat A_1$. Setting $\Delta=\omega_\nu$ and retaining only terms involving the operators $\hat a_\nu,\hat A_\nu$ gives Eq.~\eqref{eq:noncm_ham}. Since the carrier term only couples to  $\hat A_1$, it only affects the cooling of the c.m. mode, within the validity of the HPA.


\bibliography{references}

\end{document}